\documentclass[aps,prb,twocolumn,10pt,showpacs,
superscriptaddress,nofootinbib]{revtex4}
\usepackage[utf8]{inputenc}
\usepackage{amsmath}
\usepackage{amssymb}
\usepackage{color}
\usepackage{epsfig}
\usepackage{siunitx}
\usepackage{multirow}
\usepackage{tikz,graphicx}
\usepackage{blkarray}
\renewcommand{\arraystretch}{2}
\newcommand{\sy}[2][+]{\Gamma^{#1}_{#2}} 
\newcommand{\cuo}{Cu$_{\mathrm{2}}$O}
\newcommand{\bhp}{\boldsymbol{p}}
\newcommand{\bPi}{\boldsymbol{\pi}}
\newcommand{\hPi}{\pi}
\newcommand{\bn}{\boldsymbol{n}}
\newcommand{\br}{\boldsymbol{r}}
\newcommand{\be}{\boldsymbol{e}}
\newcommand{\bR}{\boldsymbol{R}}
\newcommand{\bq}{\boldsymbol{q}}
\newcommand{\bs}{\boldsymbol{S}}
\newcommand{\bI}{\boldsymbol{I}}
\newcommand{\bJ}{\boldsymbol{J}}
\newcommand{\bk}{\boldsymbol{k}}

\newcommand{\bka}{\boldsymbol{\kappa}}
\newcommand{\bK}{\boldsymbol{K}}
\newcommand{\bB}{\boldsymbol{B}}
\newcommand{\ha}{\hat{a}}
\newcommand{\bA}{\boldsymbol{A}}
\newcommand{\hn}{\hat{n}}
\newcommand{\tN}{\boldsymbol{\tilde{N}}}
\newcommand{\bv}{\boldsymbol{v}}
\newcommand{\bN}{\boldsymbol{N}}
\newcommand{\bM}{\boldsymbol{M}}
\newcommand{\bO}{\boldsymbol{O}}
\DeclareMathOperator{\Id}{Id}

\DeclareSIUnit\al{a_g}

\begin{document}
\title{Interseries transitions between Rydberg excitons in {\cuo}}

\author{Sjard Ole \surname{Kr{\"u}ger}}
\email[]{sjard.krueger@uni-rostock.de}
\author{Stefan \surname{Scheel}}
\affiliation{Institut f{\"u}r Physik, Universit{\"a}t Rostock, 
Albert-Einstein-Stra{\ss}e 23-24, D-18059 Rostock, Germany}

\date{\today}

\begin{abstract}
We investigate the infrared optical transitions between excitons of the yellow, 
green and blue series in the cuprous oxide {\cuo}. We show that, in many cases,
the dipole approximation is inadequate and, in particular, that it breaks down in 
yellow-blue transitions even for moderate principal quantum numbers of 
$n\approx 10$. The interband matrix elements of the transition operator needed 
for the evaluation of the excitonic transition strengths are derived from known 
as well as from fitted band parameters.
\end{abstract}

\maketitle

\section{Introduction}
Excitons, bound states of electrons and holes in semiconductors, have first 
been postulated by Frenkel \cite{frenkel1931} in the limit of strongly bound 
systems, and Wannier \cite{wannier1937} in the weakly bound limit. For the 
Wannier excitons, the crystal mostly acts as a dielectric background, and the 
excitonic states show remarkable similarities to those of the hydrogen atom. 
The first observation of the Wannier excitons succeeded in the yellow series of 
{\cuo} in the 1950s \cite{gross1956} reaching up to principal
quantum numbers of $n=8$. Recently, this limit has been pushed up to $n=25$ 
\cite{kazimierczuk2014} and orbital quantum numbers of $\ell=5$ 
\cite{thewes2015}, revealing an almost perfect Rydberg series. The 
nonparabolicity of the valence band does, however, induce a systematic 
deviation from the Rydberg series, which can be cast into quantum defects 
$\delta_{n,\ell}$ \cite{schoene2016_prb,schoene2016_jpb} approaching constant 
values for large $n$ similar to alkali Rydberg atoms.

Since the first observation of these excitonic Rydberg states, they have 
attracted considerable attention due to their exaggerated properties such as 
their large real-space extensions, comparatively long lifetimes and huge 
polarisabilities. The latter is responsible for the Rydberg blockade phenomenon 
already observed early on \cite{kazimierczuk2014} due to the dipole-dipole 
interaction of the yellow Rydberg excitons \cite{walther2018}, but it also
contributes to a strong interaction with the electron-hole plasma 
\cite{heckotter2018}. Further investigations have focussed on the influence 
of the valence-band structure on the excitonic quantum defects 
\cite{schoene2016_jpb,schweiner2016impact}, the interaction with phonons and 
photons \cite{stolz2018}, the influence of electric and magnetic fields 
\cite{schoene2016_prb,schweiner2017magnetoexcitons} as well as the possibility 
to observe giant-dipole excitons in crossed electromagnetic fields 
\cite{kurz2017}. Moreover, the level statistics 
\cite{assmann2016,schweiner2017magnetoexcitons2}
has been investigated showing the breaking of all anti-unitary symmetries. 

\begin{figure}
\includegraphics[width=0.9\columnwidth]{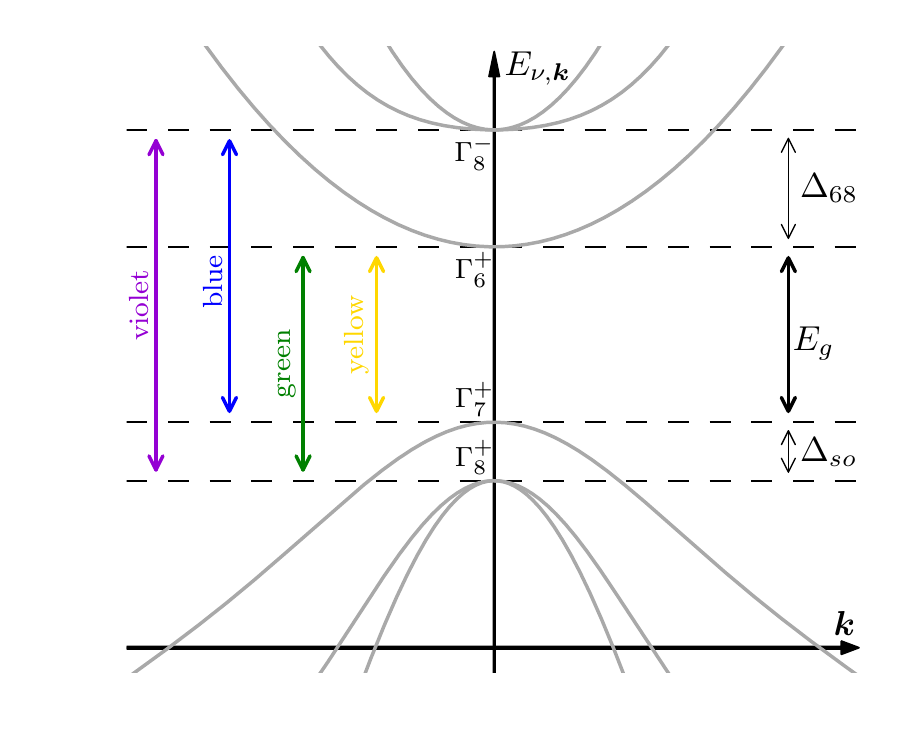}
\caption{Schematic band structure of {\cuo} in the vicinity of the 
$\Gamma$-point and the main band gap.\label{fig:band-schematic}}
\end{figure}

In addition to the yellow exciton series, there are three more excitonic 
series in {\cuo} (see Fig.~\ref{fig:band-schematic}) that have been found as 
early as the 1950s and 1960s \cite{nikitine1959,gross1962}. Transitions between 
the ground states of the yellow and blue series have been observed as 
polaritonic beating \cite{schmutzler2013}, and intraseries transitions within 
the yellow series have been probed \cite{jorger2003}. More recently, proposals
have been put forward to use the yellow intraseries transitions
for the implementation of tunable excitonic masers \cite{ziemkiewicz2018, ziemkiewicz2019}. Furthermore,
the photoluminescence of the excitons in {\cuo} has been measured, including both the blue and violet
series \cite{takahata2018.1} and the superradiance-to-polariton crossover of the blue
$1S$ state has been investigated in dependence of the crystal thickness\cite{takahata2018.2}.
As the Rydberg excitons  show coherent features already in single-photon absorption
\cite{grunwald2016}, the pertinent question is how to exploit these in coherent
manipulation schemes such as EIT-based protocols for single-photon generation \cite{khazali2017} or 
to generate giant optical nonlinearities \cite{walther2018b}. The first step in 
this direction is to identify suitable dipole-allowed transitions that are also 
easily accessible experimentally. As the excitonic Rydberg energies of {\cuo} are
very low compared to atomic systems, intraseries transitions are inconveniently 
located in the far infrared.  However, transitions between different exciton series,
i.e. interseries transitions, could be exploited. In particular, transitions
between Rydberg  states of different exciton series become accessible with
near-infrared light.

\begin{figure}
\includegraphics[width=\columnwidth]{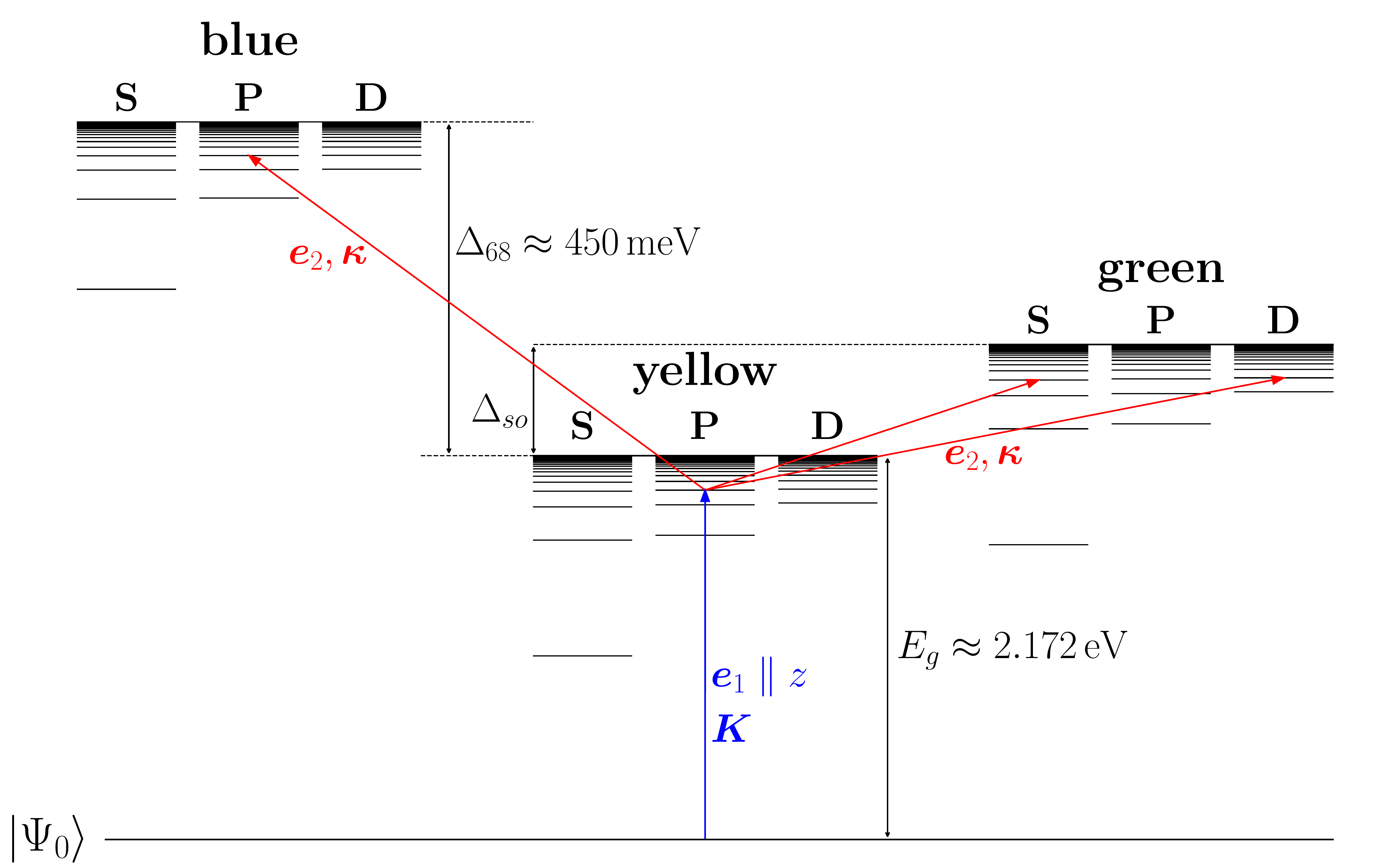}
\caption{Schematic term diagram of the excitonic states in the yellow, green 
and blue series of {\cuo}. In the proposed scheme, the $z$-component of a yellow
$\sy[-]{4}$ $P$-exciton is excited via a $z$-polarised laser beam with wave 
vector $\bK$, and the transitions into the green and blue series are probed by a
second laser beam with polarisation $\be_2$ and wave vector $\bka$. 
$|\Psi_0\rangle$ denotes the crystal vacuum. \label{fig:term-schema}}
\end{figure}

In this article, we compute the infrared transition strengths from the yellow 
$P$-excitons into the green and blue excitons as shown in 
Fig.~\ref{fig:term-schema}. These transition strengths are directly
proportional to both the transition rate per exciton and the line intensities
of the corresponding transitions. For transitions between Rydberg states of different 
series, the dipole approximation breaks down as the states can reach real-space 
extensions up to $\si{\micro\meter}$ while the separation in energy approaches 
a finite value ($\Delta_{so}=\SI{131}{\milli\eV}$ from yellow to green and 
$\Delta_{68}\approx\SI{450}{\milli\eV}$ from yellow to blue, see also 
Tab.~\ref{tab:mat-prop}). Therefore, the spatial extent of the excitons becomes 
comparable to the wavelength, in which case the dipole approximation is known to 
break down. We will first calculate the transition strengths in dipole 
approximation and subsequently assess the influence of the breakdown of the 
dipole approximation for a representative set of transitions.

\begin{table}
\def\arraystretch{1.5}
\caption{Material properties of {\cuo} used in this work.\label{tab:mat-prop}}
\begin{ruledtabular}
\begin{tabular}{r  l l | r l l}
$B_{6,8}$&$0.342\,\hbar\pi\si{\per\al}$& \multirow{3}{*}{this work} &$A_1$&$-1.76$& 
\multirow{6}{*}{\cite{schoene2016_prb}}\\
$A_6$&$2.44$&&$A_2$&$4.519$&\\
$A_8$&$3.99$&&$A_3$&$-2.201$& \\ \cline{1-3}
$E_g$&$2.17208~\si{\eV}$&\cite{kazimierczuk2014} &$B_1$&$0.02$&\\ 
\cline{1-3}
$\Delta_{68}$ & $\SI{450}{\milli\eV}$ & \cite{schmutzler2013}& $B_2$&$-0.022$& \\ \cline{1-3}
$\Delta_{so}$&$131~\si{\milli\eV}$ &\cite{gross1956}&$B_3$&$-0.202$& 
\\ \hline
$\si{\al}$ &$\SI{0.427}{\nano\meter}$ &\cite{werner1982} &  $\varepsilon_s$ & $7.5$ &\cite{carabatos1968} \\\hline
$F$ & $-0.43$ & \cite{schweiner2017magnetoexcitons} (see App. \ref{app:a}) &$\varepsilon_{\infty}$&$6.46$& \cite{kavoulakis1997}
\end{tabular}
\end{ruledtabular}
\end{table}

This article is structured as follows. In Sec.~\ref{sec:preliminaries} the 
general theory of excitonic interseries transitions is layed out both with and 
without the dipole approximation. Section~\ref{sec:interband-matrixelements} 
contains the derivation of the inter-band matrix elements of the light-matter 
coupling operator that are needed for the evaluation of the transition 
strengths. Sections~\ref{sec:yellow-green} and \ref{sec:yellow-blue} 
contain the derivation and discussion of the transition strengths from
the yellow series to the green and blue series, respectively, followed by an 
outlook on future work as well as a discussion of the results in 
Sec.~\ref{sec:outlook}. Computational details such as the chosen basis states 
and the conduction-band Hamiltonian have been delegated to the Appendix.

\section{Exciton states and transition operators}\label{sec:preliminaries}

In this section, the general structure of the transition matrix-elements of
excitonic interseries transitions will be derived. The excitons are comprised
of a bound pair of an electron in the conduction band and a hole in the valence
band. Their quantum states can be written as
\begin{equation}
\left|\Psi_{\tau, \bK}^{c,v}\right\rangle = \sum\limits_{\bk} \phi_{\tau, 
\bK}(\bk)~ \ha_{c,\bk + \alpha \bK}^{\dagger}\,\ha_{v,\bk - \beta\bK} 
|\Psi_0\rangle.
\label{eq:exciton-state}
\end{equation}
Here, $\tau=\{n,\ell,m, c, v\}$ labels the quantum numbers of the internal 
excitonic state with the envelope function $\phi_{\tau,\bK}(\bk)$, $c$ and $v$ 
denote the conduction and valence band with fermionic creation (annihilation) 
operators $\ha_{c,\bq}^{\dagger}$ ($\ha_{v,\bq}$), and $|\Psi_0\rangle$ is
the crystal vacuum state. $\bk$ and $\bK$ denote the relative and 
center-of-mass (COM) momenta of the electron and hole with the relative masses 
$\alpha=m_e/M$, $\beta = m_h/M$ and $M=m_e + m_h$. Mass estimates based on the 
band parameters are listed in Tab.~\ref{tab:exciton-prop}. 

\begin{table}
\def\arraystretch{1.5}
\caption{Estimates of the properties of the excitonic series, derived from the 
values in Tab.~\ref{tab:mat-prop} and Eq.~(\ref{eq:cb-params}). The 
experimental Rydberg energies of the green and blue series are based on only a 
few low-$n$ states and thus not very reliable.
\label{tab:exciton-prop}}

\begin{ruledtabular}
\begin{tabular}{l| c c | c c c }
 & $m_h$ & $m_e$ & $a_B/\si{\nano\meter}$ & $Ry^{*}/\si{\milli\eV}$ & 
$Ry^{*}_{\text{exp}}/\si{\milli\eV}$\\ \hline
yellow & $\frac{-m_0}{A_1-2B_1} $ &  $\frac{m_0}{A_6- 
\frac{4B_{6,8}^2}{3 m_0 \Delta_{68}}}$ & 
$1.12$ & $86.07$ & $86.04$ 
\cite{schoene2016_jpb}\\
green & $\frac{-m_0}{A_1+B_1}$ & $\frac{m_0}{A_6- 
\frac{4B_{6,8}^2}{3 m_0 \Delta_{68}}}$ & $1.09$ & $87.94$ & 
$139$ \cite{itoh1975}\\
blue & $\frac{-m_0}{A_1-2B_1}$ & $\frac{m_0}{A_8 + 
\frac{2B_{6,8}^2}{3 m_0 \Delta_{68}}}$ & $2.58$ & $37.22$ & 
$46$ \cite{gross1962}
\end{tabular}
\end{ruledtabular}
\end{table}

Note that Eq.~(\ref{eq:exciton-state}) is expressed in terms of valence-band 
electrons, and the actual hole momentum is $-\bk+\beta\bK$.  Here and in the 
following we will always assume that for the COM momenta of interest
$\phi_{\tau, \bK}(\bk) \approx  \phi_{\tau, 0}(\bk)=  \phi_{\tau}(\bk)$ holds. 
It should also be noted here that, in general, the excitonic states in 
Eq.~(\ref{eq:exciton-state}) are mixed due to the broken rotational symmetry of 
the crystal. A better approximation to the actual excitonic
states would be the use of properly symmetrised basis states as done in 
Ref.~\cite{Waters1980} and discussed in Appendix~\ref{app:symbas}. We will 
continue first the analysis in terms of the basis states in 
Eq.~(\ref{eq:exciton-state}) as the symmetrised states are linear combinations
of them, and later use the symmetrised states from Sec.~\ref{sec:yellow-green} 
onwards.

The full one-electron crystal Hamiltonian is given by 
\begin{equation}
H = \frac{\bhp^2}{2m_0} + V(r) + 
\frac{\hbar}{4\,m_0^2\,c^2}\left(\bs\times\nabla V\right)\cdot \bhp
\end{equation}
where $V(r)$ is the crystal-periodic potential containing both the effective 
interaction with the crystal ions and the other electrons, and $\bs$ denotes 
the Pauli matrices. The minimal substitution yields a light-matter coupling 
operator of the form
\begin{equation}
\frac{e\bA(\br)}{m_0}\cdot\left( \bhp  + 
\frac{\hbar}{4\,m_0\,c^2}\left(\bs\times\nabla V\right)\right) = 
\frac{e\bA(\br)}{m_0}\cdot\bPi
\end{equation}
where the Coulomb gauge has been used and the diamagnetic term has been 
ignored. Even if the components of $\bA(\br)$ do not correspond to plane waves,
they can be decomposed into them by a Fourier transformation. We will therefore 
assume the transition operator to be of the form $e^{-i\bka\br}\,\bPi$, which 
can be expressed in second quantisation as
\begin{equation}
e^{-i\bka\br}\,\bPi = \sum\limits_{\nu, \nu'} \sum\limits_{\bk} \langle 
\nu,\bk|e^{-i\bka\br}\,\bPi |\nu', \bk+\bka\rangle \, 
\ha_{\nu,\bk}^{\dagger}\,\ha_{\nu',\bk+\bka}\,,
\end{equation}
where $\nu$ and $\nu'$ sum over the bands of interest and the pseudo-momentum 
conservation has been used. Umklapp processes are being ignored due to the
small magnitude of the wave vectors of interest.

The matrix element 
$\langle\Psi_{\tau,\bK}^{c,v}|e^{-i\bka\br}\,\bPi|\Psi_{\tau',\bK'}^{c,v'}
\rangle$ with $v\ne v'$ can be evaluated as
\begin{gather*}
 \left\langle\Psi_{\tau, \bK}^{c,v}\right|e^{-i\bka\br}\,\bPi\left|\Psi_{\tau', 
\bK'}^{c,v'}\right\rangle \\
 = \sum\limits_{\nu, \nu'} \sum\limits_{\bq} \langle 
\nu,\bq|e^{-i\bka\br}\,\bPi 
 |\nu', \bq+\bka\rangle \, \sum\limits_{\bk,\bk'} \phi_{\tau'}(\bk')
 \phi_{\tau}^{\dagger}(\bk) \\
 \times\left\langle\Psi_0\right|\ha_{v,\bk - \beta\bK}^{\dagger}\,\ha_{c,\bk + 
\alpha \bK} \,\ha_{\nu,\bq}^{\dagger} \\ \times
\ha_{\nu',\bq+\bka} \ha_{c,\bk' + \alpha' \bK'}^{\dagger}\, 
\ha_{v',\bk'- \beta'\bK'} \left|\Psi_0\right\rangle.
\end{gather*}
This corresponds to a change of valence band by the hole, i.e. a transition 
from the yellow to the green series (see Fig.~\ref{fig:band-schematic}).
Using the fact that $v\ne v'$ implies that only those terms contribute in which
$\nu=v'$ and $\nu' = v$, and applying the anti-commutator rules
$[\ha^{\dagger}_{\nu,\bq},\ha_{\nu',\bq'}]_{+} = 
\delta_{\nu,\nu'}\,\delta_{\bq,\bq'}$ as well as
$[\ha_{\nu,\bq},\ha_{\nu',\bq'}]_{+} = 
[\ha^{\dagger}_{\nu,\bq},\ha^{\dagger}_{\nu',\bq'}]_{+} = 0$ gives
\begin{gather}
 \left\langle\Psi_{\tau, \bK}^{c,v}\right|e^{-i\bka\br}\,\bPi\left|\Psi_{\tau', 
\bK'}^{c,v'}\right\rangle \nonumber \\
 = -\sum\limits_{\bq} \langle v',\bq|e^{-i\bka\br}
 \,\bPi |v, \bq+\bka\rangle \sum\limits_{\bk,\bk'} \phi_{\tau'}(\bk')~ 
\phi_{\tau}^{\dagger}(\bk) \nonumber\\
\times\left\langle\Psi_0\right|\ha_{v,\bk-\beta\bK}^{\dagger}\,\ha_{v,\bq+\bka} 
\,\ha_{v',\bq}^{\dagger}\,\ha_{v',\bk' - \beta'\bK'} \nonumber \\ \times
\left(\delta_{\bk'+\alpha'\bK',\bk+\alpha\bK}-\ha_{c,\bk'+\alpha'\bK'}^{\dagger}
 \,\ha_{c,\bk + \alpha \bK}\right) 
\left|\Psi_0\right\rangle.\label{eq:transition-moment1}
\end{gather}
 
The matrix element of the vacuum state in Eq.~(\ref{eq:transition-moment1}) can 
only be nonzero if the electrons are created in the same states
from which they had been annihilated. In this case, the products of creation 
and annihilation operator can be rewritten as the number operator
$\ha_{\nu,\bq}^{\dagger}\ha_{\nu,\bq} = \hn_{\nu,\bq}$. The matrix element can 
then be evaluated by observing that, in the crystal vacuum state, all
valence bands are completely filled while the conduction bands are empty,
\begin{gather}
 \left\langle\Psi_{\tau, \bK}^{c,v}\right|e^{-i\bka\br}\,\bPi\left|\Psi_{\tau', 
\bK'}^{c,v'}\right\rangle \nonumber\\
 =  - \delta_{\bK'-\bka,\bK}\,\sum\limits_{\bk} \,\phi_{\tau}^{\dagger}(\bk + 
\beta\bK'+\alpha\bka) \phi_{\tau'}(\bk + \beta'\bK') \nonumber \\
\times \langle v',\bk|e^{-i\bka\br}\,\bPi |v, \bk+\bka\rangle.
\end{gather}
It will be more convenient for the subsequent analysis to reformulate this
equation in terms of valence-band holes instead of electrons which implies
exchanging $\langle v',\bk|e^{-i\bka\br}\,\bPi |v, \bk+\bka\rangle$ for
$\langle v,-\bk-\bka|e^{-i\bka\br}\,\bPi |v', -\bk\rangle$ and therefore
\begin{gather}
 \left\langle\Psi_{\tau, \bK}^{c,v}\right|e^{-i\bka\br}\,\bPi\left|\Psi_{\tau', 
\bK'}^{c,v'}\right\rangle \nonumber\\
 =   -\delta_{\bK'-\bka,\bK}\,\sum\limits_{\bk} \,\phi_{\tau}^{\dagger}(\bk + 
\beta\bK'+\alpha\bka) \phi_{\tau'}(\bk + \beta'\bK') \nonumber \\
\times \langle v,-\bk-\bka|e^{-i\bka\br}\,\bPi |v', -\bk\rangle.
\label{eq:transition-moment}
\end{gather}

The equivalent matrix element for a change of the conduction band (i.e. from 
the yellow to the blue series) is
\begin{gather}
\left\langle\Psi_{\tau, \bK}^{c,v}\right|e^{-i\bka\br}\,\bPi\left|\Psi_{\tau', 
\bK'}^{c',v}\right\rangle  \nonumber \\
= \delta_{\bK,\bK'-\bka}\,\sum\limits_{\bk} \phi_{\tau}^{\dagger}(\bk-\alpha 
\bK) \phi_{\tau'}(\bk- \alpha'\bK + \beta'\bka) \nonumber \\ 
\times \langle c,\bk|e^{-i\bka\br}\,\bPi |c',\bk+\bka\rangle.
\label{eq:transition-moment-conduction-band}
\end{gather}
Single-photon transitions with a change of both valence and conduction bands
(say, from the yellow to the violet series) are forbidden to all orders, as
these are two-particle transitions which require at least two photons.
In the dipole approximation (ignoring the COM momenta of both excitonic states 
and the photon $\bK = \bK'=\bka =\mathbf{0}$), Eq.~(\ref{eq:transition-moment}) 
reduces to
\begin{gather}
 \left\langle\Psi_{\tau}^{c,v\phantom{'}}\right|\bPi\left|\Psi_{\tau'}^{c,v'}\right\rangle 
 = -\sum\limits_{\bk}~\phi_{\tau}^{\dagger}(\bk)~\phi_{\tau'}(\bk) 
 \langle v,-\bk |\bPi |v', -\bk \rangle.
 \label{eq:transition-moment-dipole}
\end{gather}

The difference between Eqs.~(\ref{eq:transition-moment}) and 
(\ref{eq:transition-moment-dipole}) consists of two effects:
\begin{itemize}
 \item the non-vertical (i.e. non-dipolar) transitions between the pure Bloch 
states and
 \item the relative displacement of the envelope functions by 
$(\beta-\beta')\bK' + \alpha\bka$ which corresponds to a non-dipolar transition 
between the envelope functions.
\end{itemize}
For states with allowed dipole transitions, the first correction is expected to 
be only weak, while the second one will significantly lower the transition 
strengths for states with large principal quantum numbers $n$. For transitions 
between Rydberg states of different series, both $|\bK|$ and $|\bka|$ become 
constant while the momentum-space extension decreases $\propto n^{-1}$. This
implies that the overlap of the envelope functions will vanish and the mentioned 
approximation is only valid as long as the momentum-space extension of at least 
one of the excitonic states is much larger than $(\beta-\beta')\bK'+\alpha\bka$. 
Or, to phrase it differently, the real-space extension of at least one of the 
states has to be much smaller than the transition wavelength in order for the 
dipole approximation to be valid.

The momentum-space displacements for the yellow-green and yellow-blue 
transitions are given in Tab.~\ref{tab:mom-space-disp}. The smaller values for 
the yellow-green transitions result partially from the fact that our estimate 
for the hole mass of the $\sy{8}$ valence band is very close to that of the 
$\sy{7}$ valence band giving $\beta-\beta' \approx -0.0078$ and, additionally, 
from the larger $\bka$ of the yellow-blue transitions. This implies that the 
effects from the breakdown of the dipole approximation will be more prominent 
for the yellow-blue transitions. The wave numbers are given by 
$|\bk(E)|=E\sqrt{\varepsilon_{\infty}}/(\hbar c)$ with the refractive index 
$\sqrt{\varepsilon_{\infty}}$ of {\cuo}. The energy separations were 
approximated by the band gaps, i.e. by $E_g$ for $|\bK|$ as well as 
$\Delta_{so}$ and $\Delta_{68}$ for the yellow-green and yellow-blue photon 
momenta $|\bka|$, respectively.

\begin{table}
\def\arraystretch{1.5}
\caption{Estimates of the momentum-space displacements for the yellow-green and 
yellow-blue transitions as well as co-propagating and counter-propagating pump 
and probe beams.
\label{tab:mom-space-disp}}
\begin{ruledtabular}
\begin{tabular}{l| c | c}
& co-propagating & counter-propagating\\ \hline
yellow &$(\beta-\beta')|\bK'| + \alpha|\bka|$  & $(\beta-\beta')|\bK'| - 
\alpha|\bka|$ \\
-green&$ \approx 0.12\times 10^{-3}\pi/\si{\al}$& $ \approx -0.18\times 
10^{-3}\pi/\si{\al}$\\\hline
yellow& $(\alpha'-\alpha)|\bK| -\beta'|\bka| $&$(\alpha'-\alpha)|\bK| 
+\beta'|\bka| $\\
-blue & $\approx -1.96\times 10^{-3}\pi/\si{\al}$ & $\approx 
-0.82\times 10^{-3}\pi/\si{\al}$
\end{tabular}
\end{ruledtabular}
\end{table}

\section{Interband matrix elements}\label{sec:interband-matrixelements}

In order to proceed with the evaluation of 
Eqs.~(\ref{eq:transition-moment} -- \ref{eq:transition-moment-dipole}),
the interband matrix elements 
$\langle \nu,\bq|e^{-i(\bq'-\bq)\br}\bPi|\nu', \bq'\rangle$
as well as the envelope functions $\phi_{\tau}(\bq)$ are required. In this 
section, the calculation of the matrix elements will be layed out, while the 
envelope functions will be approximated by properly symmetrised hydrogenic 
functions with the parameters given in Tab.~\ref{tab:exciton-prop}.

For the transitions between the yellow and green series in {\cuo}, the relevant 
valence bands are the uppermost $\sy{7}$ band and the $\sy{8}$ band, that both 
stem from the same $\sy{5}$ band when the spin is ignored. The interband matrix 
elements can be rewritten in terms of the lattice periodic functions 
$|u_v,\bq\rangle$ via $|v,\bq\rangle=e^{i\bq\br}\,|u_v,\bq\rangle$ as
\begin{gather}
 \langle \sy{7},\sigma,-\bq-\bka|e^{-i\bka\br}\,\bPi |\sy{8},\sigma', -\bq\rangle 
\nonumber\\
 = \langle u_{\sy{7}},\sigma,-\bq-\bka|\bPi |u_{\sy{8}},\sigma', 
-\bq\rangle \nonumber\\
 - \hbar\bq\,\langle u_{\sy{7}},\sigma,-\bq-\bka|u_{\sy{8}}, \sigma', 
-\bq\rangle,
\label{eq:interband-1}
\end{gather}
where $\sigma$ and $\sigma'$ denote the substates of the irreducible 
representations (``spin''). This can be evaluated in perturbation theory using
\begin{equation}
|u_v,\bq\rangle \approx |u_v,0\rangle + 
\frac{\hbar}{m_0}\sum\limits_n\frac{|u_n,0\rangle
\langle u_n,0|\bq\cdot\bPi|u_v,0\rangle}{E_v(0)-E_n(0)}.
\end{equation}
The term $\propto \hbar\bq$ in Eq.~(\ref{eq:interband-1}) vanishes for 
$\bka=\mathbf{0}$, and its lowest-order term is proportional to
$q^2\kappa$. As both the $q$ and $\kappa$ of interest are small 
compared to the size of the Brillouin zone, this term will be ignored.
The remaining term in Eq.~(\ref{eq:interband-1}) gives
\begin{gather}
 \langle u_{\sy{7}},\sigma,-\bq-\bka|\bPi |u_{\sy{8}},\sigma', -\bq\rangle 
\nonumber\\
= \frac{\hbar}{m_0}\sum\limits_n\frac{
\langle u_{\sy{7}},\sigma,0|-(\bq+\bka)\cdot\bPi|u_n,0\rangle
\langle u_n,0|\bPi|u_{\sy{8}},\sigma',0\rangle}{E_{\sy{7}}(0)-E_n(0)}\nonumber\\
+ \frac{\hbar}{m_0}\sum\limits_n\frac{
\langle u_{\sy{7}},\sigma,0|\bPi|u_n,0\rangle
\langle u_n,0|-\bq\cdot\bPi|u_{\sy{8}},\sigma',0\rangle}
{E_{\sy{8}}(0)-E_n(0)} \nonumber\\
= -\langle u_{\sy{7}},\sigma,\bq|\bPi |u_{\sy{8}},\sigma',\bq\rangle \nonumber\\
- \frac{\hbar}{m_0}\sum\limits_n\frac{
\langle u_{\sy{7}},\sigma,0|\bka\cdot\bPi|u_n,0\rangle
\langle u_n,0|\bPi|u_{\sy{8}},\sigma',0\rangle}{E_{\sy{7}}(0)-E_n(0)}.
\label{eq:band-transition1}
\end{gather}
The term proportional to $\bka$ corresponds to higher-order transitions between 
the Bloch states and can induce quadrupole transitions between excitonic 
states, while the first term is responsible for the dipole transitions. 

Assuming that 
\begin{equation}
E_{\sy{7}}(0)-E_n(0)\approx E_{\sy{8}}(0)-E_n(0)\label{eq:energy-denom-approx}
\end{equation}
for all intermediate states $|u_n,0\rangle$,  Eq.~(\ref{eq:band-transition1}) 
can be rewritten as
\begin{gather}
 \langle u_{\sy{7}},\sigma,-\bq-\bka|\bPi |u_{\sy{8}},\sigma',-\bq\rangle
\nonumber\\
=-\frac{m_0}{\hbar}\nabla_{\bq + 
\frac{\bka}{2}}\,\left[\mathcal{H}\left(\bq + 
\frac{\bka}{2}\right)\right]_{\sigma, \sigma'}- F\,\hbar\tN_{\sigma, 
\sigma'}\cdot\bka.
\label{eq:band-transition2}
\end{gather}
Here, $F\approx- 0.43$ denotes a magnetic band parameter, and the matrices 
$\tN_{\sigma, \sigma'}\in\mathbb{C}^{3\times 3}$ can be derived from 
group-theoretical considerations (see Appendix~\ref{app:a}). In addition,
$\mathcal{H}(\bq)$ denotes the Suzuki-Hensel Hamiltonian \cite{suzuki1974}
\begin{gather}
\mathcal{H}(\bq) = -\frac{\Delta_{so}}{3}\,\bI\cdot\bs 
+ \frac{\hbar^2}{2m_0} \bigg\{ \left[A_1 + B_1\bI\cdot\bs\right] \bq^2 
\nonumber \\
+ \left[A_2 \left(I_x^2-\frac{1}{3}\bI^2\right) + B_2 \left(I_x\sigma_x - 
\frac{1}{3}\bI\cdot\bs\right)\right] q_x^2 + \mathrm{c.p.} \nonumber\\
+ \left[A_3 \left(I_xI_y + I_yI_x\right) + B_3 \left(I_x\sigma_y + 
I_y\sigma_x\right)\right] \{q_x,q_y\} + \mathrm{c.p.} \bigg\},
\label{eq:suzuki-hensel}
\end{gather}
where $\bI$ is the vector of the spin-1 angular-momentum matrices, $\bs$ the 
vector of the Pauli matrices and $\{q_x,q_y\} = (q_x q_y + q_yq_x)/2$
the symmetric product. Furthermore, c.p. stands for cyclic permutation,  
$\Delta_{so}$ is the spin-orbit splitting of the valence band, and the $A_i$ and 
$B_i$ are band-structure parameters whose values can be found in 
Tab.~\ref{tab:mat-prop}.

The approximation in Eq.~(\ref{eq:energy-denom-approx}) is justified as the 
intermediate states can only have $\sy[-]{7}$ or $\sy[-]{8}$ symmetry, and the 
next band of such symmetry is removed by about $20\,\Delta_{so}$ from the 
valence bands. A more in-depth discussion of Eq.~(\ref{eq:band-transition2})
as well as the matrices $\tN_{\sigma, \sigma'}$ can be found in 
Appendix~\ref{app:a}.

In this manner, the interband matrix elements are easily accessible, once the 
band structure parameters are known. For the valence bands of {\cuo}, good fits
to the Suzuki-Hensel Hamiltonian are available \cite{schoene2016_prb}. However, 
one has to keep in mind that the representation of the Hamiltonian in terms of 
$\bI$ and $\bs$ implies a basis 
$\sy[-]{4}\otimes\sy{6}=\sy[-]{6}\oplus\sy[-]{8}$. The valence band states in 
{\cuo}, however, have the symmetry
$\sy{5}\otimes\sy{6} = (\sy[-]{2}\otimes\sy[-]{4})\otimes\sy{6} = 
\sy{7}\oplus\sy{8}$. Therefore, one has to be careful when assigning the band 
states to the rows and columns of the Hamiltonian, and one cannot simply 
assign them by their eigenvalues of the operators $\bJ^2$ and $J_z$ (where 
$\bJ=\bI+\bs/2$).

\begin{figure}
\includegraphics[width=\columnwidth]{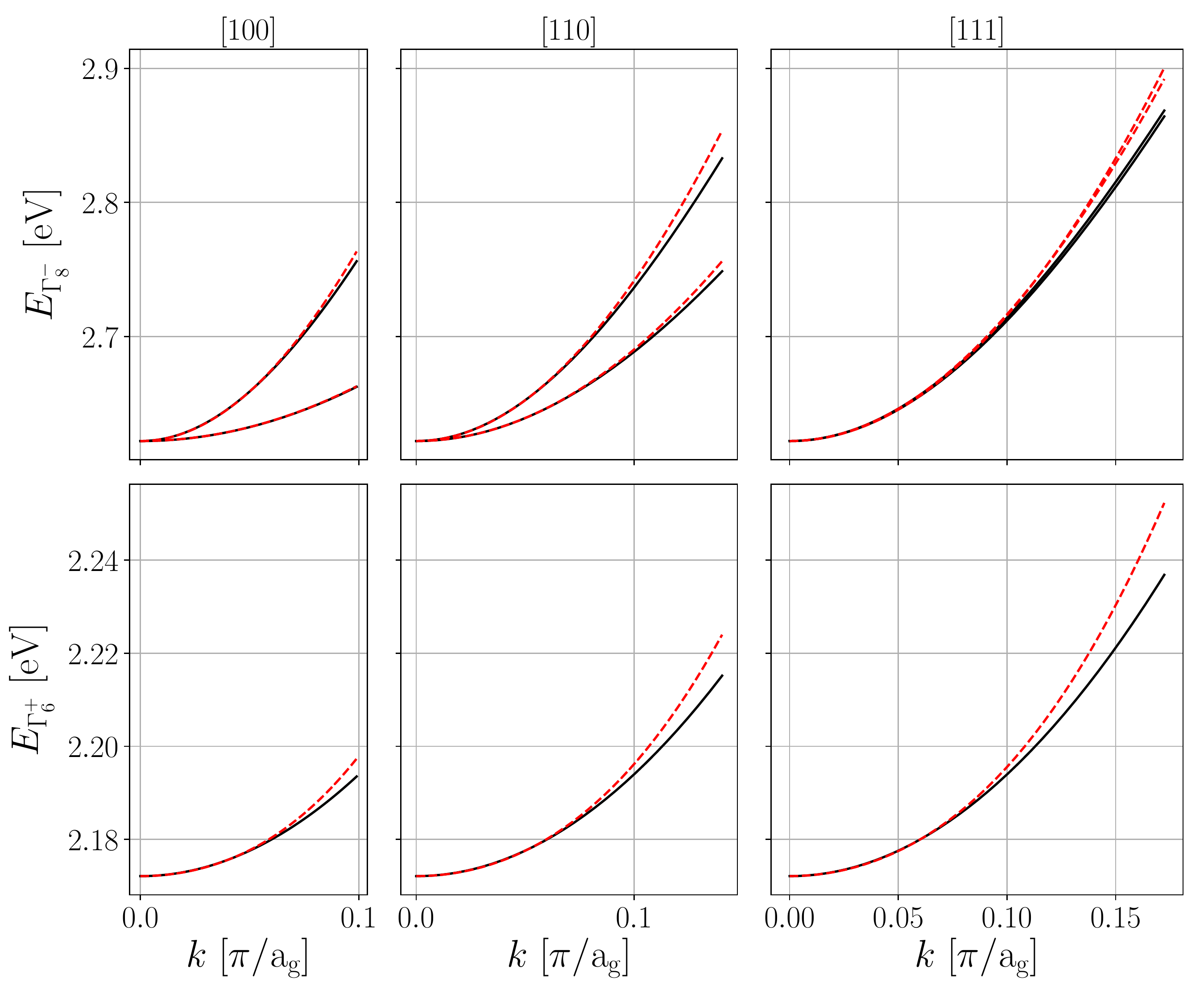}
\caption{Conduction band structure of {\cuo}, comparing results from spin-DFT 
calculations (black solid lines) with fits with the parameters in 
Eq.~(\ref{eq:cb-params}) (red dashed lines).
\label{fig:cb-structure}}
\end{figure}

For the transitions between the yellow and the blue series, the interband 
matrix element takes the form
\begin{equation}
\langle \sy{6},\sigma,\bq |e^{-i\bka\br}\,\bPi|\sy{8},\sigma',\bq+\bka\rangle 
= B_{6,8} \bR_{\sigma,\sigma'},
\end{equation}
where $B_{6,8}$ is the relevant band structure parameter and $\bR$ is given in 
Eq.~(\ref{eq:matrices-conduction68-S}). We fitted the Hamiltonian derived in 
Appendix~\ref{app:y-b-ham} to spin-DFT calculations in the vicinity of the 
$\Gamma$-point \cite{french2008}, yielding  
\begin{gather}
A_6=2.44\,\quad A_8= 3.99\,\quad A_8' =  -1.94,\nonumber\\
A_8''= 1.25 \quad\text{and}\quad B_{6,8}= 0.342 
\,\frac{\hbar\pi}{\si{\al}}.\label{eq:cb-params}
\end{gather}
The resulting band structure is shown in Fig.~\ref{fig:cb-structure}, with 
excellent agreement to the DFT calculations. Using the effective electron
masses as defined in Tab.~\ref{tab:exciton-prop} yields $m_e=0.99\,m_0$ for 
the $\sy{6}$ conduction band which agrees with the experimental values
\cite{hodby1976} as well as $m_e = 0.21\, m_0$ for the $\sy[-]{8}$ conduction
band.

\section{Transitions between the yellow and green series in {\cuo}}
\label{sec:yellow-green}

In this section, the transition matrix elements for transitions from the
yellow $P$-excitons to the green $S$- and $D$-excitons will be calculated. We 
will first evaluate them in the dipole approximation of 
Eq.~(\ref{eq:transition-moment-dipole}) and subsequently drop the dipole 
approximation for a representative set of states.

\begin{figure}
\includegraphics[width=\columnwidth]{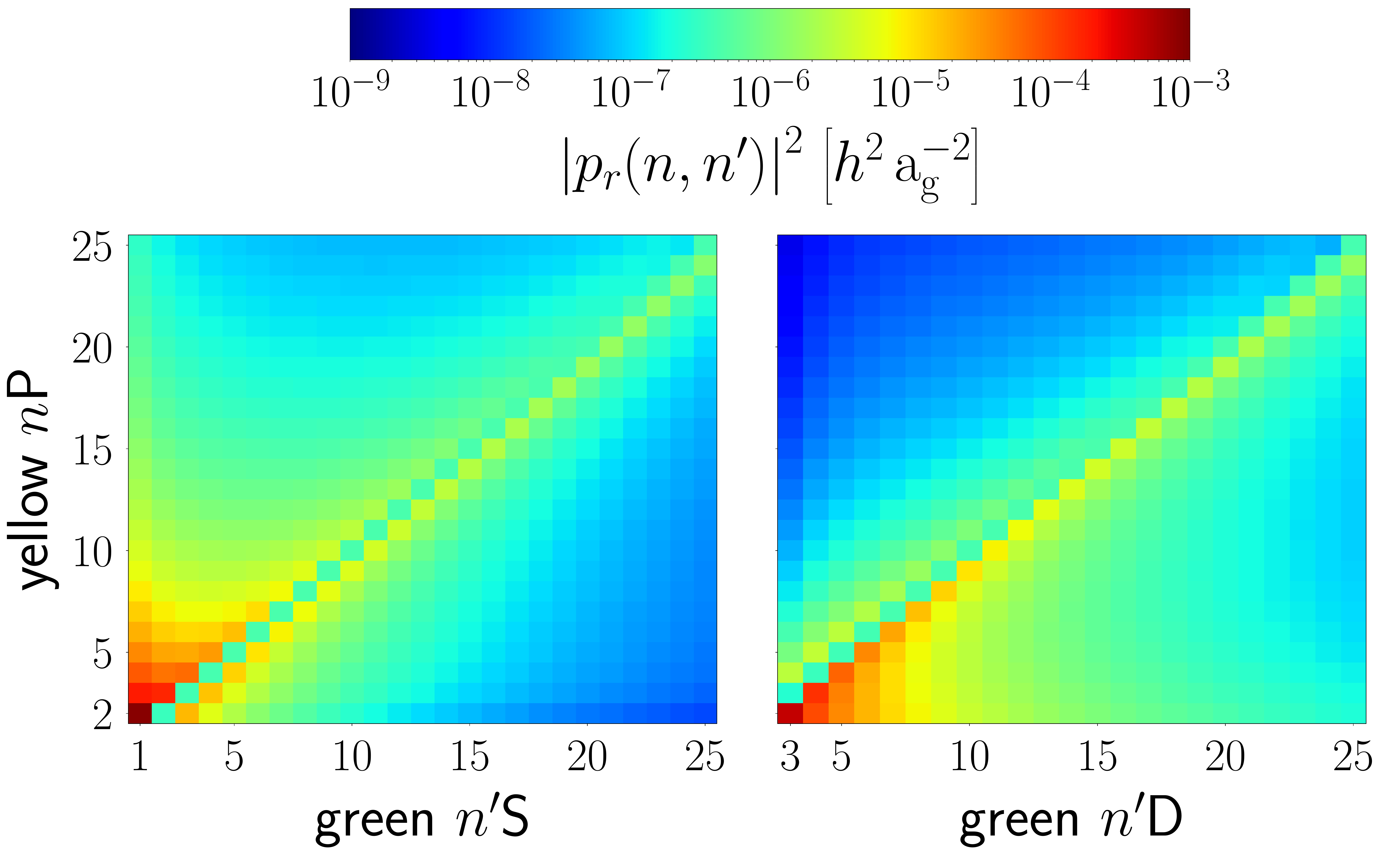}
\caption{Radial coefficients as given in Eq.~(\ref{eq:radial-coeffs}). The 
radial wave functions used are hydrogenic functions with Bohr
radii given in Tab.~\ref{tab:exciton-prop}.\label{fig:radial-coeffs}}
\end{figure}

Using the symmetrised basis functions given in Appendix~\ref{app:symbas} with 
hydrogen-like radial wave functions as approximations to the excitonic states, 
the transition matrix elements in dipole approximation can now be calculated. 
Focussing on the transitions from the $z$-component of the yellow $\sy[-]{4}$ 
$P$-exciton (without loss of generality), the transition strength to the green 
$S$- and $D$-states can be expressed as 
\begin{equation}
 \sum_i \left|\left\langle YP_z^{\sy[-]{4}},n\right|\hPi_{x_j}
 \left|G\ell^{\sy{\xi}}_i,n'\right\rangle\right|^2 
 = C_{x_j}^{\sy{\xi}} |p^{\ell}_r(n,n')|^2\label{eq:yellow-green-dipole}
\end{equation}
where the sum over $i$ runs over the substates of the green 
$\ell\sy{\xi}$-manifold, $\ell$ denotes the angular momentum (i.e $S$ or $D$) 
and $\sy{\xi}$ the irreducible representations into which the green $S$- and 
$D$-excitons can be decomposed. The $C_{\ell,x_j}^{\sy{\xi}}$ are 
listed in Tab.~\ref{tab:ang-coeff} and depend only on the spins, the angular
momenta as well as the band-structure parameters $A_2-B_2$ and $A_3-B_3$.
Furthermore,  
\begin{equation}
p^{\ell}_r(n,n') = \hbar \int\limits_0^{\infty} dk\,k^3 
\tilde{\phi}_{YP,n}^{\dagger}(k)\,\tilde{\phi}_{G\ell,n'}(k)\label{eq:radial-coeffs}
\end{equation}
are the radial matrix elements of $\hbar k$ that are shown in 
Fig.~\ref{fig:radial-coeffs}. The representation of Eq.~\ref{eq:yellow-green-dipole} as 
well as the coefficients can be retrieved by multiplying the symmetrised basis functions
in Appendix~\ref{app:symbas} with the corresponding transition matrix 
Eq.~(\ref{eq:transition-matrix-spinor-space}) and integrating over the angular 
coordinates.

In Eq.~(\ref{eq:radial-coeffs}), the sum over $\bk$ has been converted
to an integral via $\sum_{\bk}\,\approx \Omega/(2\pi)^3\,\int d^3\bk$ with
$\Omega$ the crystal volume. The radial envelope functions 
$\tilde{\phi}_{\tau}(\bk) = \sqrt{\Omega}/(2\pi)^{3/2}\,\phi_{\tau}(\bk)$ are
normalised w.r.t. the integral $\int d^3\bk$.

\begin{table}
\def\arraystretch{1.5}
\caption{Angular coefficients for dipole transitions into the green $S$- and 
$D$-excitons.\label{tab:ang-coeff}}
\begin{ruledtabular}
\begin{tabular}{|r c|c c c|}
 \multicolumn{2}{|@{\hspace{0.5em}}c|}{\multirow{2}{*}{$C_{\ell,x_i}^{\sy{\xi}}$ 
}} &\multicolumn{3}{c@{\hspace{0.5em}}|}{polarisation $x_i$ 
 ({\color{red}{$\boldsymbol{e}_2$}})}\\
 \multicolumn{2}{|@{\hspace{0.5em}}c|}{} & $x$ & $y$ & $z$ \\\hline
 \multirow{3}{*}{\shortstack[c]{green\\ $S$}}& $1\,\sy{3}~~$ & 
$~~0$ & $0$ & $0.11$\\
 & $1\,\sy{4}~~$ & $~~0.85$ & $0.85$ & $0$\\
 & $1\,\sy{5}~~$ & $~~0.52$ & $0.52$ & $0$\\
 & $\Sigma~~$ & $~~1.37$ & $1.37$ & $0.11$ \\\hline
 \multirow{5}{*}{\shortstack[c]{green\\ $D$}}& $2\,\sy{1}~~$ & 
$~~0$ & $0$ & $0.47$\\
 & $2\,\sy{2}~~$ & $~~0$ & $0$& $0$\\
 & $3\,\sy{3}~~$ & $~~0$ & $0$ & $0.87$\\
 & $5\,\sy{4}~~$ & $~~0.41$ & $0.41$ & $0$\\
 & $5\,\sy{5}~~$ & $~~0.44$ & $0.44$ & $0$\\
 & $\Sigma~~$ & $~~0.85$ & $0.85$ & $1.34$ \\\hline
\end{tabular}
\end{ruledtabular}
\end{table}

In order to include the multipole corrections contained in 
Eq.~(\ref{eq:transition-moment}), we first observe that for interband 
transitions between bands of the same parity, the leading order of the interband 
matrix element can be expressed as
\begin{equation}
 \langle v,-\bq-\bka|e^{-i\bka\br}\,\bPi |v', -\bq \rangle 
 = -\hbar\bM \cdot \bq - \hbar\bN\cdot\bka
\end{equation}
with some state-dependent matrices $\bM, \bN \in\mathbb{C}^{3\times 3}$. In 
this approximation, Eq.~(\ref{eq:transition-moment}) can be rewritten as
\begin{gather}
 \left\langle\Psi_{\tau, \bK}^{c,v}\right|e^{-i\bka\br}\,\bPi
 \left|\Psi_{\tau', 
\bK+\bka}^{c,v'}\right\rangle \nonumber \\
 = i\,\hbar\, \bM \cdot \bO(\tau, \tau', (\beta-\beta')\bK+ \alpha'\bka) 
\nonumber\\
 - \hbar\left[\bN\cdot\bka - \beta'\bM\cdot(\bK+\bka)\right]W(\tau, \tau', 
(\beta-\beta')\bK+ \alpha'\bka)
\label{eq:yellow-green-nondipole}
\end{gather}
where 
\begin{gather}
\bO(\tau, \tau', \bq) = \int d^3\br~e^{i\bq\cdot\br}\, 
\phi_{\tau}^{\dagger}(\br)\,\nabla_{\br}\, \phi_{\tau'}(\br)\label{eq:bO}
\end{gather}
and
\begin{gather}
W(\tau, \tau',\bq) = \int d^3\br~e^{i\bq\cdot\br}\, 
\phi_{\tau}^{\dagger}(\br)\, \phi_{\tau'}(\br)\label{eq:W}
\end{gather}
with the real-space (i.e. Fourier transformed) envelope functions 
$\phi_{\tau}(\br)$. The integrals $W(\tau, \tau',\bq)$ are identical to those
appearing in the calculation of the phonon scattering of Rydberg excitons 
\cite{stolz2018}.
Table~\ref{tab:rel-transition-strength} lists the transition strengths relative 
to the dipole approximation for some representative transitions from the 
yellow $P$-excitons to the green $\sy{3}$-excitons. For counter-propagating beams and 
a principal quantum number of $\sim 20$, the error introduced by the 
dipole approximation can reach almost $25\%$. 

\begin{table}
\def\arraystretch{1.5}
\caption{Relative transition strengths from $n_y\sy[-]{4}$ $P$-excitons to all 
green $n_g \sy{3}$-excitons with both pump and probe beam $z$-polarised and 
propagating along the $y$-axis.\label{tab:rel-transition-strength}}
\begin{ruledtabular}
\begin{tabular}{c c| c  c | c c}
 $n_y$ & $n_g$ & \multicolumn{2}{c|}{co-propagating} & \multicolumn{2}{c}{counter-propagating}\\
  & & $S\,\sy{3}$&$D\,\sy{3}$ &$S\,\sy{3}$ &$D\,\sy{3}$ \\\hline
 $19$ & $20$ & $1.04$ & $0.96$ & $0.76$& $0.83$ \\
 $14$ & $15$ & $1.02$& $0.99$  & $0.90$ &$0.94$\\
 $8$ & $9$ &$1.01$ & $1.00$ & $0.98$ & $0.99$ 
\end{tabular}
\end{ruledtabular}
\end{table}

\section{Transitions between the yellow and blue series in {\cuo}}
\label{sec:yellow-blue}

We will now proceed by repeating the analysis of the previous section
for transitions from the yellow to the blue $P$-excitons. In particular, we 
will show that the breakdown of the dipole approximation allows 
transitions to certain blue $S$-excitons with transition strengths comparable 
to those from yellow $P$ to blue $P$-excitons.

\begin{figure}
\includegraphics[width=0.9\columnwidth]{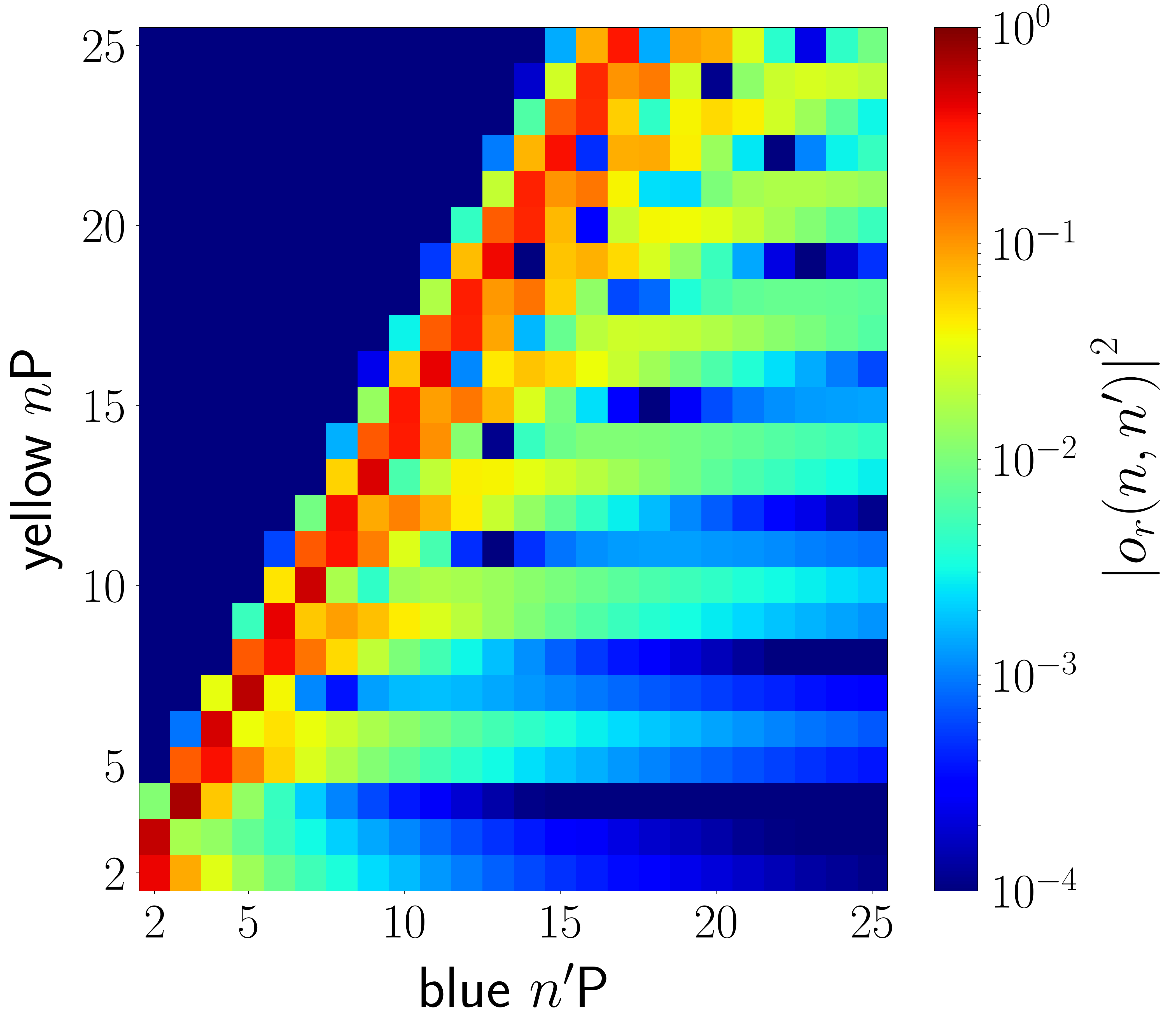}
\caption{Radial coefficients as given in Eq.~(\ref{eq:radial-coeffs-2}) for the 
transitions between the yellow and blue $P$-excitons. The radial wave functions 
used are hydrogenic functions with Bohr radii given in 
Tab.~\ref{tab:exciton-prop}.\label{fig:radial-coeffs-2}}
\end{figure}

The effective mass, averaged spatially as well as over light and heavy holes, 
of an electron in the $\sy[-]{8}$ conduction band derived from the fits is 
$0.21\,m_0$. Note that some of this mass derives from the coupling to the 
$\sy{6}$ conduction band. Hence, this value is not the same as the inverse of 
$A_8$ in Eq.~(\ref{eq:cb-params}). As the interband matrix elements are 
constant to leading order in $\bq$, the only allowed transitions in the dipole 
approximation are those between states of the same $\ell$. The dipolar
transition strengths between yellow $nP$ and blue $n'P$ excitons can therefore 
be written as
\begin{gather}
  \sum_i \left|\left\langle YP_z^{\sy[-]{4}},n\right|\hPi_{x_j}
\left|BP^{\sy{\xi}}_i, n'\right\rangle\right|^2 \nonumber \\
  = \left|B_{6,8}\right|^2 D_{P, x_j}^{\sy{\xi}} |o_r(n,n')|^2
\end{gather}
with the band parameter $B_{6,8}$ (see Appendix~\ref{app:y-b-ham}) and the 
radial overlap integral
\begin{equation}
o_r(n,n') = \int\limits_0^{\infty} dk~k^2\,
\tilde{\phi}_{YP,n}^{\dagger}(k)\,\tilde{\phi}_{BP,n'}(k).
\label{eq:radial-coeffs-2}
\end{equation}
The blue $P$ states have symmetry $\sy[-]{4}\otimes\sy[-]{8}\otimes\sy{7} = 
\sy{1}\oplus\sy{2}\oplus 2\sy{3}\oplus 3\sy{4} \oplus 3\sy{5}$, and the 
corresponding coefficients $D_{P, x_j}^{\sy{\xi}}$ are given in 
Tab.~\ref{tab:ang-coeff-yb} with the squared overlap integrals shown in 
Fig.~\ref{fig:radial-coeffs-2}.

\begin{table}
\def\arraystretch{1.5}
\caption{Angular coefficients for dipole transitions into the blue 
$P$-excitons.\label{tab:ang-coeff-yb}}
\begin{ruledtabular}
\begin{tabular}{|r c|c c c|}
 \multicolumn{2}{|@{\hspace{0.5em}}c|}{\multirow{2}{*}{$D_{P, x_i}^{\sy{\xi}}$ }} 
 &\multicolumn{3}{c@{\hspace{0.5em}}|}{polarisation $x_i$ 
({\color{red}{$\boldsymbol{e}_2$}})}\\
 \multicolumn{2}{|@{\hspace{0.5em}}c|}{} & $x$ & $y$ & $z$ \\ \hline
 \multirow{3}{*}{\shortstack[c]{blue\\ P}}& $1\,\sy{1}~~$ & $~~0$ & $0$ & 
$1/3$\\
  & $1\,\sy{2}~~$ & $~~0$ & $0$ & $0$\\
   & $2\,\sy{3}~~$ & $~~0$ & $0$ & $1/3$\\
 & $3\,\sy{4}~~$ & $~~5/12$ & $5/12$ & $0$\\
 & $3\,\sy{5}~~$ & $~~1/4$ & $1/4$ & $0$\\
 & $\Sigma~~$ & $~~2/3$ & $2/3$ & $2/3$ \\\hline
\end{tabular}
\end{ruledtabular}
\end{table}

\begin{figure}
\includegraphics[width=\columnwidth]{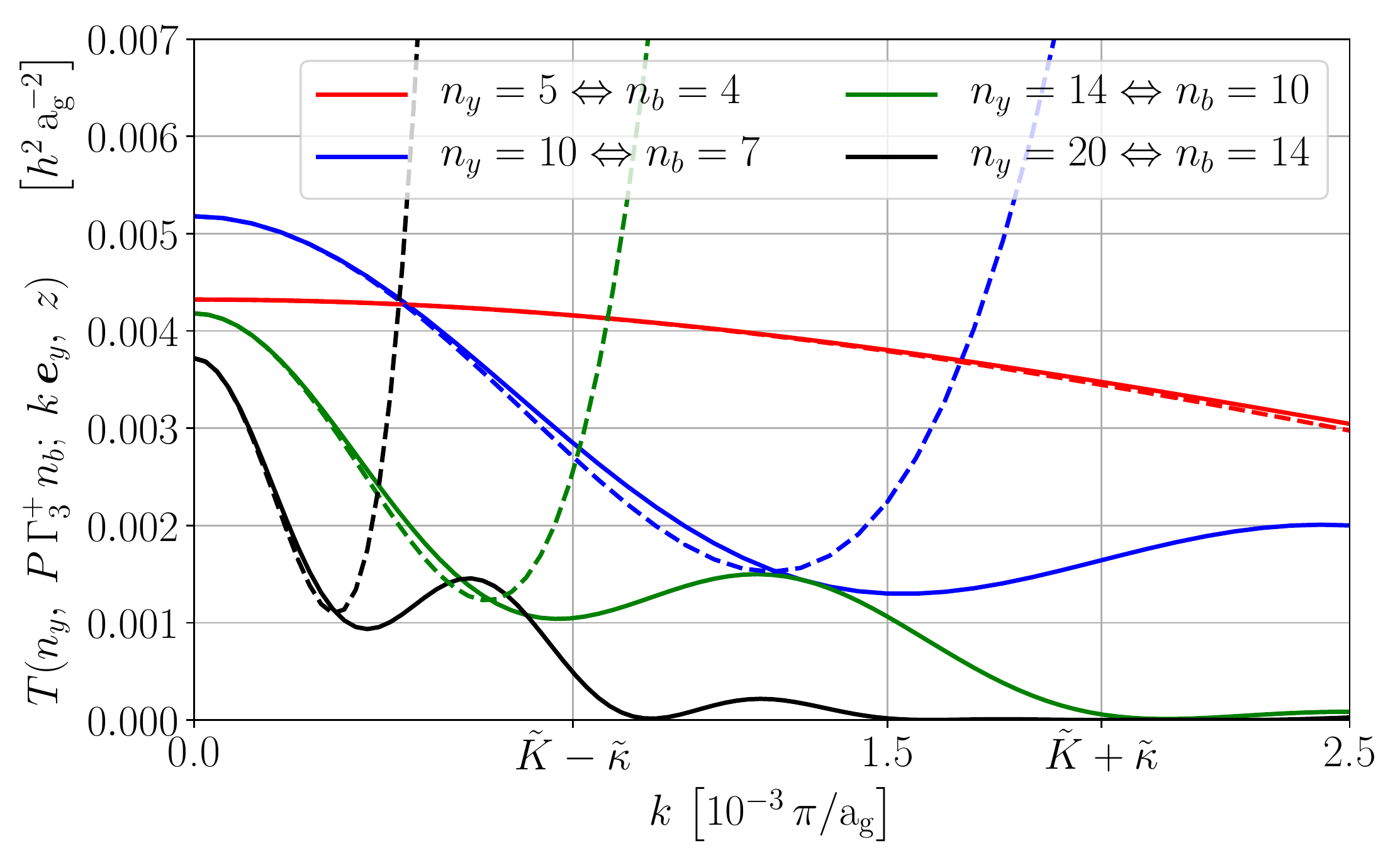}
\caption{Transitions from the $z$-component of the yellow $n_y \sy[-]{4}$ 
$P$-state to all blue $n_b \sy{3}$ $P$-states via a $z$-polarised probe beam.
The dashed lines indicate the sum of the dipole 
and octupole contributions.
\label{fig:beyond-dipole-blue}}
\end{figure}

For a constant interband transition matrix element
$\langle c,\bk|e^{-i\bka\br}\,\bPi|c',\bk+\bka \rangle\approx \langle c,0|\,\bPi|c',0 \rangle$,
Eq.~(\ref{eq:transition-moment-conduction-band}) can be 
rewritten as
\begin{gather}
\left\langle\Psi_{\tau, \bK}^{c,v}\right|e^{-i\bka\br}\,\bPi
\left|\Psi_{\tau', \bK+\bka}^{c',v}\right\rangle =
\langle c,0|\bPi |c', 0\rangle \nonumber\\
\times W(\tau, \tau',(\alpha'-\alpha)\bK - \beta'\bka).
\label{eq:yellow-blue-nondipole}
\end{gather}
The corrections to transitions from the $z$-component of the yellow 
$n_y\,\sy[-]{4}$ $P$-exciton to all blue $n_b\,\sy{3}$ $P$-states are
shown in Fig.~\ref{fig:beyond-dipole-blue}. The quantity
\begin{gather}
 T(n_y,~ P\,\sy{3}\, n_b;~ \bk,~ x_j) = |B_{6,8}|^2\sum\limits_j \\
 \times\left|\left\langle YP_z^{\sy[-]{4}}\, n_y\right| R_{x_j}^{\dagger}
\otimes \bar{W}(YP\,n_y,\, BP\, n_b,\,\bk)\left|BP_j^{\sy{3}}\,n_b\right\rangle\right|^2\nonumber
\end{gather}
shown there coincides with the exact transition strength for
$\bk = (\alpha'-\alpha)\bK - \beta'\bka$ and with the 
dipole approximation for $\bk=0$. Here, $j$ sums over all four substates of the
the two $\sy{3}$ representations appearing in the decomposition of the blue
$P$-excitons (see Tab.~\ref{tab:ang-coeff-yb}),
$\bar{W}(\tau, \tau', \bk)\in\mathbb{C}^{(2\ell+1)\times (2\ell'+1)}$
denotes the matrix containing the $W(\tau, \tau', \bk)$ for all combinations
of magnetic quantum numbers, the $R_{x_j}$
are the interband coefficient matrices given in
Eq.~(\ref{eq:matrices-conduction68-S}), and 
$|BP_j^{\sy{3}}\,n_b\rangle$ as well as $|YP_z^{\sy[-]{4}}\, n_y\rangle$ 
are the symmetrised basis states of App.~\ref{app:symbas}.

For the given relative electron masses in {\cuo}, the scaled COM wavenumber
of the yellow exciton is 
$\tilde{K} = (\alpha-\alpha')| \bK| =1.39\times 10^{-3}\pi\si{\per\al}$, and
$\tilde{\kappa} = \beta'|\bka|=0.57\times 10^{-3} \pi\si{\per\al}$ is the 
photon wavenumber scaled by the relative hole mass of the blue exciton.
These values have been calculated from the band gaps of the yellow and blue 
series and should be accurate to within $1\%$ even for the 
$n_y=5\Leftrightarrow n_b=4$ transition. The sum $\tilde{K}+\tilde{\kappa}$
appears for co-propagating beams and the difference $\tilde{K}-\tilde{\kappa}$ 
for counter-propagating beams (see Tab.~\ref{tab:mom-space-disp}).

\begin{figure}
\includegraphics[width=\columnwidth]{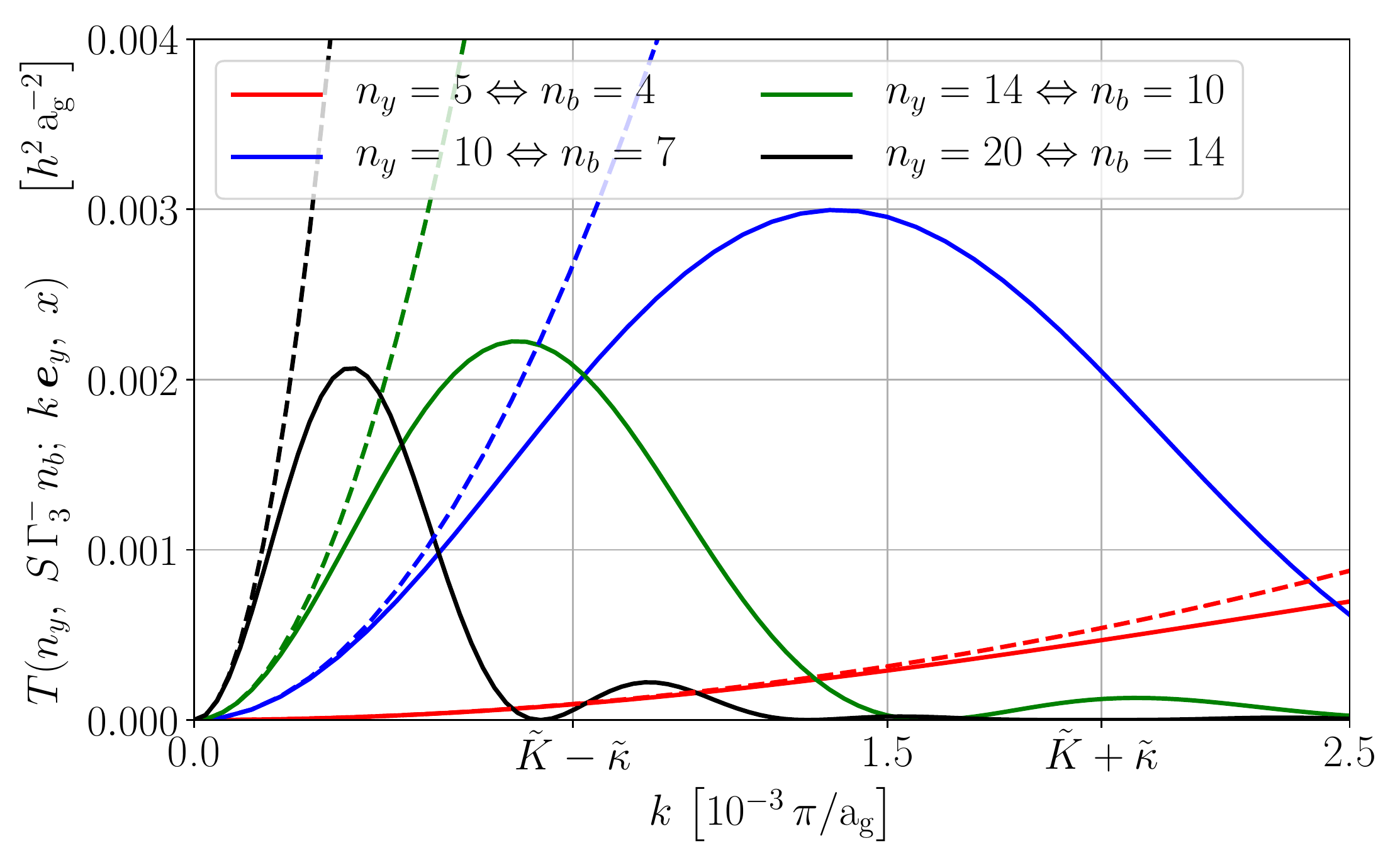}
\caption{Same as Fig.~\ref{fig:beyond-dipole-blue} for the transition to 
the blue $\sy[-]{3}$ $S$-excitons via $x$-polarised probe beams.
Dashed lines indicate the quadrupole approximation.
\label{fig:beyond-dipole-blue-s}}
\end{figure}

There is a prominent effect from the abandonment of the dipole approximation 
even for principal quantum numbers $n_y\approx 5$. In 
Fig.~\ref{fig:beyond-dipole-blue}, the dashed lines show the sum of dipole 
and octupole contributions only, which deviates strongly from the exact result 
(solid lines) for larger momentum transfer $k$. However, the deviation of
the exact result from the low-order multipole expansion is least for
counter-propagating beams. In Fig.~\ref{fig:beyond-dipole-blue-s}, we show
the transition strengths from the yellow $\sy[-]{4}\,P$-excitons to the blue
$\sy[-]{3}\,S$-excitons which are usually dipole-forbidden. Hence, the first
non-vanishing contribution is that of a quadrupole interaction. Note that the
magnitudes of the transition matrix elements are of the same order of magnitude
as in Fig.~\ref{fig:beyond-dipole-blue}.

\section{Discussion and outlook}
\label{sec:outlook}

We have presented detailed calculations for the transition strengths of the 
excitonic interseries transitions in {\cuo}, going well beyond the dipole 
approximation. Those transitions can either be between the yellow and green 
series, where the hole changes the valence band, or between the yellow and blue 
series with an accompanying change of the conduction band. The symmetry 
properties of the respective bands imply that dipole transitions are only 
allowed for $|\Delta\ell|=1$ (yellow-green) or $\Delta\ell=0$ (yellow-blue), 
respectively. 

Transitions between Rydberg excitons of different series are located at 
wavelengths that are comparable with the size of the Rydberg wavefunctions. 
Hence, the dipole approximation is no longer valid, and the inclusion of
multipoles of all orders is necessary. We have shown that, already for 
relatively low principal quantum numbers, the deviation from the dipole 
approximation cannot be neglected. In particular, dipole-forbidden transitions 
such as from yellow $P$-excitons to blue $S$-excitons become allowed to the 
extent that their transition matrix elements are of similar size to those of 
dipole-allowed transitions. These results pave the way to construct protocols 
for coherent manipulation of Rydberg excitons.

So far, our calculations are based on the exciton parameters collected in 
Tab.~\ref{tab:exciton-prop} which were derived from band-structure parameters of
{\cuo} and show discrepancies to the experimental data. To improve on these, a
detailed theory of the green \cite{schweiner2017even, alvermann2018} and blue 
Rydberg excitons, including the coupling to the yellow continuum states, would 
be necessary. Furthermore, both the $\sy{8}$ valence band and the $\sy[-]{8}$ 
conduction band have an anisotropic mass at the $\Gamma$-point that will result 
in a strong coupling of different angular momenta even for Rydberg states and a 
corresponding redistribution of the transition strengths over more states than 
we have taken into account so far. To the best of our knowledge, no experiments
have been performed, from which the excitonic interseries transition strengths
in {\cuo} could be quantified.

Proposals have been put forward to use spatially modulated light fields such as 
orbital angular momentum (OAM) light in order to tune the selection rules of 
excitonic transitions \cite{konzelmann2019}. Realising such transitions is, 
however, experimentally challenging as the real-space extension of both initial 
and final states has to be comparable to the characteristic length scale of the 
modulation, or the transversal position inside the beam has to be controlled 
very precisely \cite{schmiegelow2016, afanasev2018}.
Excitonic interseries transitions, on the other hand, have the advantage that 
both initial and final states can in principle become arbitrarily large while 
the energy separation, and hence the wavelength of the coupling beam,
approaches a finite value. This work paves the way towards the evaluation of 
such transition matrix elements which we leave for a future publication.

\begin{acknowledgments}
We would like to thank Dr. Martin French for providing us with the 
high-resolution SDFT calculations of the conduction bands in the vicinity of 
the $\Gamma$-point. We gratefully acknowledge support by the DFG SPP 
1929 GiRyd.
\end{acknowledgments}

\appendix
\section{Interband matrix elements}
\label{app:a}

The Suzuki--Hensel Hamiltonian in the $\sy{7}$--$\sy{8}$ cross space 
represents the perturbation theoretical $\bk\cdot\bPi$-Hamiltonian of the 
form \cite{kp-voon}
\begin{gather}
  \left[\mathcal{H}_{\bq}+ \mathcal{H}_{\bB}\right]_{\sigma,\sigma'}  
  = \frac{\hbar^2}{2m_0^2}\sum\limits_n \Delta_n^{-1} \nonumber\\
  \times\langle u_{\sy{7}},\sigma,0|\bq\cdot\bPi|u_n,0\rangle
  \langle u_n,0|\bq\cdot\bPi|u_{\sy{8}},\sigma',0\rangle + 
\mathcal{O}\left(\bq^4\right)
\end{gather}
where 
\begin{equation}
\Delta_n^{-1} = \frac{1}{E_{\sy{8}}(0)-E_n(0)} + \frac{1}{E_{\sy{7}}(0)-E_n(0)},
\end{equation}
$\mathcal{H}_{\bq}$ denotes only the $\bq$-dependent part
of $\mathcal{H}(\bq)$, and the magnetic field $\bB$ is given
by the commutator \cite{luttinger1956,suzuki1974}
\begin{equation}
\bq\times\bq = \frac{e\bB}{i\hbar}.\label{eq:k-commutator}
\end{equation}
In this notation, $\sigma=\{\pm1/2\}$ and 
$\sigma'=\{\pm1/2,\pm3/2\}$ denote the valence band ``spin''.
The representations $\sy{7}$ and $\sy{8}$ are left out for notational
convenience but are nonetheless implied. For a vanishing magnetic field
one obtains
\begin{equation}
 \left[\mathcal{H}_{\bq}\right]_{\sigma,\sigma'}  
 =\frac{\hbar\bq}{2m_0}\cdot
 \langle u_{\sy{7}},\sigma,\bq|\bPi|u_{\sy{8}},\sigma', \bq\rangle
\end{equation}
as can be seen from Eq.~(\ref{eq:band-transition1}). This can be rewritten as
\begin{equation}
\left[\mathcal{H}_{\bq}\right]_{\sigma,\sigma'}  
=\frac{\hbar^2}{2m_0}\bq\cdot\boldsymbol{M}_{\sigma, \sigma'}\cdot\bq
\end{equation}
and
\begin{equation}
\langle u_{\sy{7}},\sigma,\bq|\bPi |u_{\sy{8}},\sigma', \bq\rangle = \hbar\boldsymbol{M}_{\sigma, \sigma'}\cdot\bq 
\end{equation}
with some state-dependent matrix 
$\boldsymbol{M}_{\sigma,\sigma'}\in\mathbb{C}^{3\times 3}$. For a symmetric 
matrix $\boldsymbol{M}_{\sigma, \sigma'}$ it follows that 
$\nabla_{\bq}(\bq\cdot\boldsymbol{M}_{\sigma, \sigma'}\cdot\bq) = 
2\boldsymbol{M}_{\sigma, \sigma'}\cdot\bq$, which would imply 
\begin{gather}
\langle u_{\sy{7}},\sigma,\bq|\bPi |u_{\sy{8}},\sigma', \bq\rangle 
= m_0\hbar^{-1}\nabla_{\bq}\,\left[\mathcal{H}_{\bq}\right]_{\sigma,\sigma'}\nonumber\\
= m_0\hbar^{-1}\nabla_{\bq}\,\left[\mathcal{H}(\bq)\right]_{\sigma,\sigma'}.
\end{gather}

All that is left to show is hence, that the $\boldsymbol{M}$ are indeed 
symmetric matrices. Their components can be written as 
\begin{gather}
 \boldsymbol{M}_{\sigma,\sigma'}^{x_i, x_j} \nonumber\\
 = \frac{1}{m_0}\sum\limits_n\frac{\langle 
u_{\sy{7}},\sigma,0|\hPi_{x_j}|u_n,0\rangle
\langle u_n,0|\hPi_{x_i}|u_{\sy{8}},\sigma',0\rangle}{E_{\sy{7}}(0)-E_n(0)} 
\nonumber \\
 + \frac{1}{m_0}\sum\limits_n\frac{\langle u_{\sy{7}},\sigma, 
0|\hPi_{x_i}|u_n,0\rangle
\langle u_n,0|\hPi_{x_j}|u_{\sy{8}},\sigma',0\rangle}{E_{\sy{8}}(0)-E_n(0)}.
\end{gather}
They are obviously symmetric w.r.t. an exchange of $x_i$ and $x_j$ under the 
approximation that the two energy denominators are equal. Again, this
approximation is reasonable as the next possible coupling states 
$|u_n,0\rangle$ (of $\sy[-]{7}$ or $\sy[-]{8}$ symmetry) are removed by about 
$20\Delta_{so}$. The components of 
$m_0\,\hbar^{-1}\nabla_{\bq}\,\mathcal{H}\left(\bq \right)$ are then given by 
$\hbar \bM_{\sigma, \sigma'}\cdot\bq$ where the $\bM$ are defined by
\begin{equation}
\bM_{\sigma, \sigma'}=~\begin{blockarray}{ccc}
\langle\sy{7},-1/2|~ &~~\langle\sy{7},\phantom{-}1/2| &\\
\begin{block}{(cc)c}
  \bM_{5,z}  &  \bM_{3,1} &~~ |\sy{8},-3/2\rangle \\
 \bM_{3,2}  +\bM_{5,xy}& \sqrt{\frac{1}{3}} \bM_{5,z}&~~ |\sy{8},-1/2\rangle \\
\sqrt{\frac{1}{3}} \bM_{5,z}^{\dagger} & -\bM_{3,2} +\bM_{5,xy} &~~ |\sy{8},\phantom{-}1/2\rangle\\
-\bM_{3,1} &  \bM_{5,z}^{\dagger}&~~ |\sy{8},\phantom{-}3/2\rangle \\
\end{blockarray}
\end{equation}
with 
\begin{equation}
 \bM_{5,z} = \frac{A_3-B_3}{\sqrt{8}}\begin{pmatrix}
        0 & 0 & 1\\
        0 & 0 & i\\
        1 & i & 0
       \end{pmatrix},
\end{equation}
\begin{equation}
 \bM_{3,1} = \frac{A_2-B_2}{\sqrt{18}}\begin{pmatrix}
        -1 & 0 & 0\\
        0 & -1 & 0\\
        0 & 0 & 2
       \end{pmatrix},
\end{equation}
\begin{equation}
 \bM_{3,2} = \frac{A_2-B_2}{\sqrt{6}}\begin{pmatrix}
        1 & 0 & 0\\
        0 & -1 & 0\\
        0 & 0 & 0
       \end{pmatrix},
\end{equation}
and
\begin{equation}
 \bM_{5,xy} = \frac{A_3-B_3}{\sqrt{6}}\begin{pmatrix}
        0 & i & 0\\
        i & 0 & 0\\
        0 & 0 & 0
       \end{pmatrix}.
\end{equation}

Under the same approximations, the $\bka$-dependent part of 
Eq.~(\ref{eq:band-transition1}) can be rewritten as 
\begin{gather}
\frac{\hbar}{m_0}\sum\limits_n\frac{
\langle u_{\sy{7}},\sigma,0|\bka\cdot\bPi|u_n,0\rangle
\langle u_n,0|\bPi|u_{\sy{8}},\sigma',0\rangle}
{E_{\sy{7}}(0)-E_n(0)} 
\nonumber\\
= \frac{m_0}{2\hbar}\nabla_{\bka}\,
\left[\mathcal{H}(\bka)\right]_{\sigma,\sigma'} + 
F\tN_{\sigma,\sigma'}\cdot\hbar\bka 
\label{eq:n-term-kappa}
\end{gather}
where $\tN_{\sigma,\sigma'}
=(\boldsymbol{N}_{\sigma,\sigma'}-\boldsymbol{N}_{\sigma,\sigma'}^T)/2$ is the 
antisymmetric part of
\begin{gather}
\boldsymbol{N}_{\sigma,\sigma'}^{x_i,x_j} = 
\nonumber\\
\frac{1}{F\,m_0}\sum\limits_n\frac{
\langle u_{\sy{7}},\sigma,0|\hPi_{x_j}|u_n,0\rangle
\langle u_n,0|\hPi_{x_i}|u_{\sy{8}},\sigma',0\rangle}
{E_{\sy{7}}(0)-E_n(0)}
\label{eq:N-matrix}
\end{gather}
and $F$ is an additional parameter. We thus arrive at
\begin{gather}
 \langle u_{\sy{7}},\sigma,\bq+\bka|\bPi |u_{\sy{8}},\sigma', \bq+\bka\rangle
\nonumber \\
=  \frac{m_0}{\hbar}\nabla_{\bq + \frac{\bka}{2}}\,
  \left[\mathcal{H}\left(\bq + \frac{\bka}{2}\right)\right]_{\sigma, \sigma'}+ 
F\,\tN_{\sigma,\sigma'}\cdot\hbar\bka.\label{eq:yellow-green-interband-me}
\end{gather}

The matrices $\tN$ for all state combinations can be calculated by the 
applying the Wigner--Eckart theorem of $O_h$ to Eq.~(\ref{eq:N-matrix}) 
and a subsequent antisymmetrisation which results in
\begin{equation}\tN_{\sigma,\sigma'}=~~~
\begin{blockarray}{ccc}
\langle\sy{7},-1/2|~ &~~\langle\sy{7},\phantom{-}1/2|& \\
\begin{block}{(cc)c}
 \sqrt{\frac{1}{3}} \tN_{4,xy}  &  \tN_{4,z}&~~|\sy{8},-3/2\rangle \\
 0 & - \tN_{4,xy}&~~|\sy{8},-1/2\rangle \\
-\tN_{4,xy}^{*} & 0 &~~|\sy{8},\phantom{-}1/2\rangle\\
\tN_{4,z} & \sqrt{\frac{1}{3}} \tN_{4,xy}^{*}&~~|\sy{8},\phantom{-}3/2\rangle \\
\end{blockarray}
\end{equation}
where the $^{*}$ denotes a complex conjugation and
\begin{equation}
 \tN_{4,xy} = \sqrt{\frac{1}{2}}\,\begin{pmatrix}
        0 & 0 & -1\\
        0 & 0 & -i\\
        1 & i & 0
       \end{pmatrix}
\end{equation}
as well as
\begin{equation}
 \tN_{4,z} = \sqrt{\frac{2}{3}}\,\begin{pmatrix}
        0 & i & 0\\
        -i & 0 & 0\\
        0 & 0 & 0
       \end{pmatrix}.
\end{equation}

Due to the antisymmetry of $\tN$, one can rewrite 
$\bka\cdot\tN_{\sigma,\sigma'}\cdot\bka = e/(i 
\hbar)\,\bv_{\sigma,\sigma'}\cdot\bB$ where
\begin{equation}
 \bv_{4,xy} = \sqrt{\frac{1}{2}}\,\begin{pmatrix}
        -i\\1\\0
       \end{pmatrix}
,\quad
 \bv_{4,z} = \sqrt{\frac{2}{3}}\,\begin{pmatrix}
        0\\0\\i
       \end{pmatrix}
\end{equation}
and the commutation relation in Eq.~(\ref{eq:k-commutator}) has been used. This
tells us that the parameter $F$ is related to the magnetic part of the valence-band
Hamiltonian which is given by (ignoring spin-orbit terms) \cite{suzuki1974}
\begin{gather}
 \mathcal{H}_{\bB} = \frac{\hbar e}{m_0}\left[ A_4 \bI + B_4 \bs \right]\cdot\bB.
\end{gather}

Inserting $[E_{\sy{7}}(0)-E_n(0)]^{-1} \approx \Delta_n^{-1}/2$ 
into Eq.~(\ref{eq:n-term-kappa}), we can thus write
\begin{gather}
  \frac{\hbar^2 F}{2m_0} \bka \cdot\tN_{\sigma,\sigma'}\cdot\bka
  = \frac{1}{2}\left[\mathcal{H}_{\bB}\right]_{\sigma,\sigma'}\nonumber\\
\Rightarrow\frac{\hbar e}{2m_0}\, F\, \frac{\bv_{\sigma,\sigma'}}{i}\,\cdot\bB  
= \frac{\hbar e}{2m_0}\,\left[ A_4 \bI_{\sigma,\sigma'} 
+ B_4 \bs_{\sigma,\sigma'} \right]\cdot\bB
\end{gather}
and find by comparison $F = -(A_4-2B_4)/\sqrt{3}$. A further comparison with the 
magnetic valence-band Hamiltonian  
\begin{equation}
  \mathcal{H}_{\bB} = \frac{\hbar e}{m_0}\left[\left(\frac{3\kappa}{2}
+\frac{g_s}{4}\right) \bI-\frac{g_s}{4} \bs \right]\cdot\bB
\end{equation}
and its parameters given in Ref.~\cite{schweiner2017magnetoexcitons} yields 
\begin{equation}
F = -\sqrt{\frac{3}{4}} \left(\kappa + \frac{g_s}{2}\right) \approx -0.43 
\end{equation}
where $g_s\approx 2$ and $\kappa\approx -0.5$ is the fourth Luttinger parameter 
\cite{luttinger1956}. Note that the spin matrices are defined differently 
there,  $\bI$ differs by a factor of $\hbar$ and $\bs_h = \hbar\bs/2$.
For the transition between Rydberg excitons of different parity, the term in 
Eq.~(\ref{eq:n-term-kappa}) will be small compared to both the transition 
strength and the correction from the displacement of the momentum-space 
envelope functions in Eq.~(\ref{eq:transition-moment}).

In the case of transitions between the yellow and blue series, the matrix 
elements of the $\sy{6}$ and $\sy[-]{8}$ conduction bands are of interest. 
Transitions between these bands are allowed at the $\Gamma$-point and the band 
Hamiltonian thus takes the form
\begin{equation}
  \left[\mathcal{H}_{\bq}\right]_{\sigma, \sigma'} 
  = \frac{\hbar}{m_0}\langle u_{\sy{6}},\sigma,0|\bq\cdot\bPi|u_{\sy[-]{8}},\sigma',0\rangle
  + \mathcal{O}\left(\bq^3\right)
\end{equation}
which is equivalent to Eq.~(\ref{eq:hamiltonian68}) and directly contains the 
sought after transition matrix elements via
\begin{gather}
 \langle u_{\sy{6}},\sigma,\bq|\bPi|u_{\sy[-]{8}},\sigma',\bq+\bka\rangle
=\frac{m_0}{\hbar}\nabla_{\bq}\left[\mathcal{H}_{\bq}\right]_{\sigma,\sigma'}\\
=\frac{m_0}{\hbar}\nabla_{\bq}\left[\mathcal{H}(\bq)\right]_{\sigma,\sigma'}
=\langle u_{\sy{6}},\sigma,0|\bPi|u_{\sy[-]{8}},\sigma',0\rangle\nonumber.
\end{gather}
The next higher-order terms would be of order $q^2$ and $q\kappa$.

\section{Symmetrised basis states}
\label{app:symbas}

Here we list the symmetrised basis states for the yellow $\sy[-]{4}$ 
$P$-excitons and the green $S$- and $D$-excitons which are equivalent to the 
one used in Ref.~\cite{Waters1980}. Note, that all following states are constructed
in terms of valence band holes. The product states of the bands forming the 
yellow series give $\sy{7}\otimes\sy{6} = \sy{2}\oplus\sy{5}$ where 
\cite{koster1963}
\begin{gather}
Y^{\sy{2}} =\sqrt{\frac{1}{2}}\left(\psi^{\sy{6}}_{1/2}\,\psi^{\sy{7}}_{-1/2} 
- \psi^{\sy{6}}_{-1/2}\,\psi^{\sy{7}}_{1/2}\right),\label{eq:waters-y-1}\\
Y^{\sy{5}}_{yz} 
=i\sqrt{\frac{1}{2}}\left(\psi^{\sy{6}}_{1/2}\,\psi^{\sy{7}}_{1/2} - 
\psi^{\sy{6}}_{-1/2}\,\psi^{\sy{7}}_{-1/2}\right),\\
Y^{\sy{5}}_{zx} 
=\sqrt{\frac{1}{2}}\left(\psi^{\sy{6}}_{1/2}\,\psi^{\sy{7}}_{1/2} + 
\psi^{\sy{6}}_{-1/2}\,\psi^{\sy{7}}_{-1/2}\right),\\
Y^{\sy{5}}_{xy} 
=-i\sqrt{\frac{1}{2}}\left(\psi^{\sy{6}}_{1/2}\,\psi^{\sy{7}}_{-1/2} + 
\psi^{\sy{6}}_{-1/2}\,\psi^{\sy{7}}_{1/2}\right).
\label{eq:waters-y-4}
\end{gather}
In the same vein, the states of the green series can be expressed as 
$\sy{8}\otimes\sy{6} = \sy{3}\oplus\sy{4}\oplus\sy{5}$
\begin{gather}
G^{\sy{3}}_1 = 
-\sqrt{\frac{1}{2}}\left(\psi^{\sy{6}}_{1/2}\,\psi^{\sy{8}}_{-1/2} + 
\psi^{\sy{6}}_{-1/2}\,\psi^{\sy{8}}_{1/2}\right),\label{eq:waters-g-1}\\
G^{\sy{3}}_2 = 
-\sqrt{\frac{1}{2}}\left(\psi^{\sy{6}}_{1/2}\,\psi^{\sy{8}}_{3/2} + 
\psi^{\sy{6}}_{-1/2}\,\psi^{\sy{8}}_{-3/2}\right),\\
G^{\sy{4}}_x = 
i\sqrt{\frac{1}{8}}\bigg(\sqrt{3}\left[\psi^{\sy{6}}_{-1/2}\,\psi^{\sy{8}}_{3/2} 
+ \psi^{\sy{6}}_{1/2}\,\psi^{\sy{8}}_{-3/2}\right]\\
-\left[\psi^{\sy{6}}_{-1/2}\,\psi^{\sy{8}}_{-1/2} + 
\psi^{\sy{6}}_{1/2}\,\psi^{\sy{8}}_{1/2}\right]\bigg),\nonumber\\
G^{\sy{4}}_y = 
\sqrt{\frac{1}{8}}\bigg(\sqrt{3}\left[\psi^{\sy{6}}_{-1/2}\,\psi^{\sy{8}}_{3/2} 
- \psi^{\sy{6}}_{1/2}\,\psi^{\sy{8}}_{-3/2}\right]\\
+\left[\psi^{\sy{6}}_{-1/2}\,\psi^{\sy{8}}_{-1/2} - 
\psi^{\sy{6}}_{1/2}\,\psi^{\sy{8}}_{1/2}\right]\bigg),\nonumber\\
G^{\sy{4}}_z = 
i\sqrt{\frac{1}{2}}\left(\psi^{\sy{6}}_{1/2}\,\psi^{\sy{8}}_{-1/2}-\psi^{\sy{6}}
_{-1/2}\,\psi^{\sy{8}}_{1/2}\right),
\end{gather}
\begin{gather}
G^{\sy{5}}_{yz} = 
-i\sqrt{\frac{1}{8}}\bigg(\left[\psi^{\sy{6}}_{-1/2}\,\psi^{\sy{8}}_{3/2} + 
\psi^{\sy{6}}_{1/2}\,\psi^{\sy{8}}_{-3/2}\right]\\
+\sqrt{3}\left[\psi^{\sy{6}}_{-1/2}\,\psi^{\sy{8}}_{-1/2} + 
\psi^{\sy{6}}_{1/2}\,\psi^{\sy{8}}_{1/2}\right]\bigg),\nonumber\\
G^{\sy{5}}_{zx} = 
\sqrt{\frac{1}{8}}\bigg(\left[\psi^{\sy{6}}_{-1/2}\,\psi^{\sy{8}}_{3/2} - 
\psi^{\sy{6}}_{1/2}\,\psi^{\sy{8}}_{-3/2}\right]\\
-\sqrt{3}\left[\psi^{\sy{6}}_{-1/2}\,\psi^{\sy{8}}_{-1/2} - 
\psi^{\sy{6}}_{1/2}\,\psi^{\sy{8}}_{1/2}\right]\bigg),\nonumber\\
G^{\sy{5}}_{xy} = 
i\sqrt{\frac{1}{2}}\left(\psi^{\sy{6}}_{1/2}\,\psi^{\sy{8}}_{3/2}-\psi^{\sy{6}}_
{-1/2}\,\psi^{\sy{8}}_{-3/2}\right).
\label{eq:waters-g-8}
\end{gather}

The valence-band transition matrix of Eq. (\ref{eq:yellow-green-interband-me})
in the 12-dimensional spinor-space of 
$\sy{6}\otimes(\sy{7}\oplus\sy{8})$ is simply
\begin{equation}
 \left[\frac{m_0}{\hbar}\nabla_{\bq + \frac{\bka}{2}}\,
 \left[\mathcal{H}\left(\bq + \frac{\bka}{2}\right)\right]_{\sigma, \sigma'}+ 
F\,\hbar\tN\cdot\bka\right]\otimes\Id_{6}
\label{eq:transition-matrix-spinor-space}
\end{equation}
where $\Id_{6}$ is the unity operator in the $\sy{6}$-Hilbert space of the 
conduction band. The yellow $\sy[-]{4}$ $P$-excitons can then be constructed 
with the cubic harmonics \cite{vdLage1947} for $\ell=1$
\begin{gather}
\xi_x^{1,\sy[-]{4}} =  \frac{Y_1^{-1}-Y_1^1}{\sqrt{2}},\quad
\xi_y^{1,\sy[-]{4}} = i\frac{Y_1^{-1}+Y_1^1}{\sqrt{2}},\,\quad
\xi_z^{1,\sy[-]{4}} = Y_1^{0}
\end{gather}
as
\begin{gather}
 YP^{\sy[-]{4}}_x = \frac{\phi}{\sqrt{2}}\,\left(i Y^{\sy{5}}_{xy}\, 
 \frac{Y^{-1}_1 + Y_1^1}{\sqrt{2}}  + Y^{\sy{5}}_{zx}\, Y^{0}_1\right),\\
 YP^{\sy[-]{4}}_y = \frac{\phi}{\sqrt{2}}\,\left(Y^{\sy{5}}_{xy}\, 
 \frac{Y^{-1}_1 - Y_1^1}{\sqrt{2}}  + Y^{\sy{5}}_{yz}\, Y^{0}_1\right),\\
 YP^{\sy[-]{4}}_z = \frac{\phi}{2}\,\left(Y^{\sy{5}}_{zx}\, 
 \left[Y^{-1}_1 - Y_1^1\right] + i Y^{\sy{5}}_{yz}\, \left[Y^{-1}_1 + 
Y_1^1\right]\right),
\end{gather}
where the $Y_{\ell}^m$ are the spherical harmonics and $\phi$ denotes the 
radial wave function which is assumed to depend only on $n$, $\ell$ and the 
excitonic series (i.e. yellow, green or blue). 

The green $S$-states can be constructed by multiplying each of the product 
states in Eqs.~(\ref{eq:waters-g-1})--(\ref{eq:waters-g-8}) by $\phi\, Y_0^0$. 
There are 40 green $D$-states in total, which will not be listed here but can 
easily be constructed with the coupling constants given in 
Ref.~\cite{koster1963} and the cubic harmonics for $\ell=2$:
\begin{gather}
 \xi_{1}^{2,\sy{3}} =  Y^{0}_{2},\quad 
 \xi_{2}^{2,\sy{3}} =  \frac{Y^{-2}_{2} + Y^{2}_{2}}{\sqrt{2}} ,\nonumber\\
\xi_{yz}^{2,\sy{5}} = i\frac{Y^{-1}_{2} + Y^{1}_{2}}{\sqrt{2}},\quad
\xi_{zx}^{2,\sy{5}} =\frac{Y^{-1}_{2} - Y^{1}_{2}}{\sqrt{2}} ,\nonumber \\
\xi_{xy}^{2,\sy{5}} =i\frac{Y^{-2}_{2} - Y^{2}_{2}}{\sqrt{2}}.
\end{gather}

The blue product states form 
$\sy[-]{8}\otimes\sy{7}=\sy[-]{3}\oplus\sy[-]{4}\oplus\sy[-]{5}$ where
\begin{gather}
B^{\sy[-]{3}}_1 = 
-\sqrt{\frac{1}{2}}\left(\psi^{\sy{7}}_{-1/2}\,\psi^{\sy[-]{8}}_{-3/2} 
+\psi^{\sy{7}}_{1/2}\,\psi^{\sy[-]{8}}_{3/2}\right),\label{eq:waters-b-1}\\
B^{\sy[-]{3}}_2 = 
\sqrt{\frac{1}{2}}\left(\psi^{\sy{7}}_{-1/2}\,\psi^{\sy[-]{8}}_{1/2} 
+ \psi^{\sy{7}}_{1/2}\,\psi^{\sy[-]{8}}_{-1/2}\right),\\
B^{\sy[-]{4}}_x = -i\sqrt{\frac{1}{8}}
\bigg(\sqrt{3}\left[\psi^{\sy{7}}_{-1/2}\,\psi^{\sy[-]{8}}_{-1/2} 
+ \psi^{\sy{7}}_{1/2}\,\psi^{\sy[-]{8}}_{1/2}\right]\nonumber\\
+\left[\psi^{\sy{7}}_{-1/2}\,\psi^{\sy[-]{8}}_{3/2} 
+ \psi^{\sy{7}}_{1/2}\,\psi^{\sy[-]{8}}_{-3/2}\right]\bigg),
\end{gather}
\begin{gather}
B^{\sy[-]{4}}_y = \sqrt{\frac{1}{8}}
\bigg(\sqrt{3}\left[\psi^{\sy{7}}_{1/2}\,\psi^{\sy[-]{8}}_{1/2} 
- \psi^{\sy{7}}_{-1/2}\,\psi^{\sy[-]{8}}_{-1/2}\right]\nonumber\\
+\left[\psi^{\sy{7}}_{-1/2}\,\psi^{\sy[-]{8}}_{3/2} 
- \psi^{\sy{7}}_{1/2}\,\psi^{\sy[-]{8}}_{-3/2}\right]\bigg),\\
B^{\sy[-]{4}}_z = 
i\sqrt{\frac{1}{2}}\left(\psi^{\sy{7}}_{1/2}\,\psi^{\sy[-]{8}}_{3/2}
-\psi^{\sy{7}}_{-1/2}\,\psi^{\sy[-]{8}}_{-3/2}\right),\\
B^{\sy[-]{5}}_{yz} = 
-i\sqrt{\frac{1}{8}}\bigg(\left[\psi^{\sy{7}}_{-1/2}\,\psi^{\sy[-]{8}}_{-1/2} 
+ \psi^{\sy{7}}_{1/2}\,\psi^{\sy[-]{8}}_{1/2}\right]\nonumber\\
-\sqrt{3}\left[\psi^{\sy{7}}_{-1/2}\,\psi^{\sy[-]{8}}_{3/2} 
+ \psi^{\sy{7}}_{1/2}\,\psi^{\sy[-]{8}}_{-3/2}\right]\bigg),
\end{gather}
\begin{gather}
B^{\sy[-]{5}}_{zx} = 
\sqrt{\frac{1}{8}}\bigg(\left[\psi^{\sy{7}}_{-1/2}\,\psi^{\sy[-]{8}}_{-1/2} 
- \psi^{\sy{7}}_{1/2}\,\psi^{\sy[-]{8}}_{1/2}\right]\nonumber\\
+\sqrt{3}\left[\psi^{\sy{7}}_{-1/2}\,\psi^{\sy[-]{8}}_{3/2} 
- \psi^{\sy{7}}_{1/2}\,\psi^{\sy[-]{8}}_{-3/2}\right]\bigg),\\
B^{\sy[-]{5}}_{xy} = 
i\sqrt{\frac{1}{2}}\left(\psi^{\sy{7}}_{1/2}\,\psi^{\sy[-]{8}}_{-1/2}
-\psi^{\sy{7}}_{-1/2}\,\psi^{\sy[-]{8}}_{1/2}\right).
\label{eq:waters-b-8}
\end{gather}
There are 24 blue $P$-states, which can be constructed in the same manner from 
these product states and the coupling coefficients given by Koster 
\cite{koster1963}. The interband matrix for transitions between the yellow and 
blue series can be expressed as $\Id_7\otimes\left[B_{6,8} \bR\right]$.

\section{Conduction-band Hamiltonian}
\label{app:y-b-ham}

In order to describe the transitions between the yellow and the blue series, a 
band Hamiltonian describing the $\sy{6}$ and $\sy[-]{8}$ conduction bands is 
needed (or at least the offdiagonal part thereof).
For the $\sy{6}$-subspace there is one free parameter for the terms of zeroth 
order in $\bq$ ($\hat{=}\sy{1}$) and one parameter for the terms of second 
order 
($\hat{=}\sy[-]{4}\otimes\sy[-]{4}=\sy{1}\oplus\sy{3}\oplus\sy{4}\oplus\sy{5}$) 
in $\bq$ as 
$\langle\sy{6}|\sy{3}|\sy{6}\rangle=\langle\sy{6}|\sy{5}|\sy{6}\rangle=0$
and the $\sy{4}$-component corresponds to the commutator of the components of 
$\bq$, which vanishes without a magnetic field. The Hamiltonian is thus
\begin{equation}
\mathcal{H}_6(\bq) =  E_6  +\frac{\hbar^2\bq^2}{2m_0}  A_6 
\end{equation}
with the free parameters $E_6$ and $A_6$.

For the $\sy[-]{8}$-subspace, the analysis is similar, only that the angular 
momentum matrices for spin $I=3/2$ 
\begin{gather}
I_x = \frac{1}{2}
\begin{pmatrix}0 &\sqrt{3} & 0 & 0\\ \sqrt{3} & 0 & 2 & 0\\ 0 & 2 & 0 & 
\sqrt{3}\\ 0 & 0 & \sqrt{3} & 0\end{pmatrix},\nonumber\\
I_y = \frac{i}{2}
\begin{pmatrix}0 &\sqrt{3} & 0 & 0\\ -\sqrt{3} & 0 & 2 & 0\\ 0 & -2 & 0 & 
\sqrt{3}\\ 0 & 0 & -\sqrt{3} & 0\end{pmatrix},\nonumber\\
I_z = \frac{1}{2}
\begin{pmatrix}-3 &0 & 0 & 0\\ 0 & -1 & 0 & 0\\ 0 & 0 & 1 & 0\\ 0 & 0 & 0 & 
3\end{pmatrix}
\end{gather}
have to be used, and that there are in total 3 parameters for the second-order 
terms (one for $\sy{1}$, $\sy{3}$ and $\sy{5}$ each). This results in the 
Hamiltonian
\begin{gather}
  \mathcal{H}_8(\bq) \nonumber\\
 =E_8 +  \frac{\hbar^2}{2m_0}\bigg\{ A_8\, \bq^2   + 
A_8'\left(\left[I_x^2-\frac{1}{3}\bI^2\right]q_x^2 + 
\text{c.p.}\right)\nonumber\\
  + A_8'' \left(\left\{I_x,I_y\right\}\left\{q_x, q_y\right\} + 
\text{c.p.}\right) \bigg\}
\end{gather}
where $\left\{I_x,I_y\right\} = (I_x\, I_y + I_y\,I_x)/2$ is the symmetrised 
product and $\text{c.p.}$ denotes the cyclically permuted terms. This is 
equivalent to the Suzuki--Hensel Hamiltonian in the subspace of the $\sy{8}$ 
valence bands.

Dipole transitions between the two conduction bands are allowed at the 
$\Gamma$-point, therefore the lowest-order term in the cross-space is linear in
$\bq$ ($\hat{=}\sy[-]{4}$). There is one free parameter, and this part of the 
Hamiltonian can be constructed from the three matrices that transform as
$\sy[-]{4}$ in the $\sy[-]{8}\otimes\sy{6}$ cross-space \cite{koster1963}
\begin{gather}
U_x = i\sqrt{\frac{1}{2}}
\begin{pmatrix}-1&0\\ 0 &-\sqrt{\frac{1}{3}}\\\sqrt{\frac{1}{3}} & 0 \\ 0 
&1\end{pmatrix},\,
U_y = \sqrt{\frac{1}{2}}
\begin{pmatrix}1&0\\ 0 &\sqrt{\frac{1}{3}}\\\sqrt{\frac{1}{3}} & 0 \\ 0 
&1\end{pmatrix}, \nonumber\\
U_z = -i\sqrt{\frac{2}{3}}
\begin{pmatrix}0&0\\ 1 & 0 \\ 0 & 1 \\ 0 & 0\end{pmatrix}.\label{eq:u-matrices-conduction68-S}
\end{gather}
The off-diagonal Hamiltonian can then be expressed via
\begin{equation}
 \mathcal{H}_{6,8}(\bq) = \frac{\hbar}{m_0} B_{6,8}\, \bR\cdot\bq
 \label{eq:hamiltonian68}
\end{equation}
where
\begin{gather}
 R_{x_i} = \begin{pmatrix}0 & U_{x_i}^{\dagger}\\U_{x_i} & 0\end{pmatrix} 
 \in\mathbb{C}^{6\times6} \label{eq:matrices-conduction68-S}
\end{gather}
and the complete Hamiltonian is simply
\begin{equation}
 \mathcal{H}(\bq) = \mathcal{H}_6(\bq)+\mathcal{H}_8(\bq) + \mathcal{H}_{6,8}(\bq).
\end{equation}
The parameter of interest for the transitions between yellow and blue series is thus $B_{6,8}$.

\section{Evaluation of the integrals}
\label{app:integrals}

The two types of overlap integrals in Eqs.~(\ref{eq:bO}) and (\ref{eq:W}) have 
been evaluated by using the plane-wave expansion 
\begin{gather}
 e^{i\bq\cdot\br}  = 4\pi\sum\limits_{\ell=0}^{\infty}\,
 \sum\limits_{m=-\ell}^{\ell}\,(-1)^m\, 
i^{\ell}\,j_{\ell}(qr)\,Y_{\ell}^{m}(\bn_{\bq})\,Y_{\ell}^{-m}(\bn_{\br})
\end{gather}
where $j_{\ell}(r)$ denotes the spherical Bessel functions. Inserting this into 
Eq.~(\ref{eq:W}) and integrating over the angular degrees of freedom yields 
\begin{gather}
W(\tau, \tau',\bq) = 4\pi\sum\limits_{\lambda=|\ell-\ell'|}^{\ell+\ell'}\,
(-1)^{m'}\, i^{\lambda}\,Y_{\lambda}^{m'-m}(\bn_{\bq})\nonumber\\
\times G(\lambda,m-m'; \ell, -m; \ell', m') \int\limits_0^{\infty} dr~r^2\,
j_{\lambda}(qr)\, f_{\tau}^{\dagger}(r)\, f_{\tau'}(r)
\end{gather}
where $f_{\tau}(r)$ is the radial part of $\phi_{\tau}(\br)$ and  
$G(\ell_1,m_1;\ell_2, m_2; \ell_3, m_3)$ denotes the Gaunt coefficients
\begin{gather}
 G(\ell_1, m_1; \ell_2, m_2; \ell_3, m_3) \nonumber \\
 = \oint d^2\bn~Y_{\ell_1}^{m_1}(\bn)Y_{\ell_2}^{m_2}(\bn)Y_{\ell_3}^{m_3}(\bn) 
\nonumber \\
 = \sqrt{\frac{(2\ell_1+1)(2\ell_2+1)(2\ell_3+1)}{4\pi}} \nonumber\\
 \times
\begin{Bmatrix}\ell_1 & \ell_2 &\ell_3\\0 &0 & 
0\end{Bmatrix}
\begin{Bmatrix}\ell_1 & \ell_2 &\ell_3\\m_1 & m_2 & m_3\end{Bmatrix}.
\nonumber
\end{gather}
The symbols in the curly brackets are the Wigner 3-j symbols. The Gaunt  
coefficients vanish if $m_1 + m_2 + m_3 \ne 0$ or $\ell_1 + \ell_2 + \ell_3$ is 
an odd number.

The other integrals can be evaluated through
\begin{widetext}
\begin{gather}
  \bO(\tau,\tau', \bq) =  - 4\pi\sum\limits_{\mu=-1}^1 
\,(-1)^{m'-\mu}\,\langle \ell'+1, m'-\mu; 1, \mu|\ell', m'\rangle\, 
\sum\limits_{\lambda = |\ell'+1-\ell|}^{\ell'+1+\ell} G(\ell, -m; \ell'+1, 
m'-\mu; \lambda, m-m'+\mu)\,\, i^{\lambda} 
\nonumber\\
\times\,Y_{\lambda}^{m'-\mu-m}(\bn_{\bq}) \,\boldsymbol{\xi}_{\mu}\, 
\left[\frac{\ell'+1}{2\ell'+1}\right]^{1/2}\int\limits_0^{\infty} 
dr~r^2\,j_{\lambda}(qr)\,f_{\tau}^{\dagger}(r)\,
\left(\frac{\partial}{\partial r} -\frac{\ell'}{r}\right)f_{\tau'}(r) 
\nonumber\\
 +4\pi \sum\limits_{\mu=-1}^1 (-1)^{m'-\mu}\,
 \langle \ell'-1, m'-\mu; 1, \mu|\ell', m'\rangle\, 
 \sum\limits_{\lambda = |\ell'-1-\ell|}^{\ell'-1+\ell}  
 G(\ell, -m; \ell'-1,m'-\mu; \lambda,m+\mu-m') \, i^{\lambda} 
 \nonumber\\
 \times\,Y_{\lambda}^{m'-\mu-m}(\bn_{\bq})\,\boldsymbol{\xi}_{\mu}
 \,\left[\frac{\ell'}{2\ell'+1}\right]^{1/2}\,
 \int\limits_0^{\infty} dr~r^2\,\,j_{\lambda}(qr)\,
f_{\tau}^{\dagger}(r)\,
\left(\frac{\partial }{\partial r} +\frac{\ell'+1}{r}\right)f_{\tau'}(r)
 \end{gather}
\end{widetext} 
where $\langle \ell_1, m1; \ell_2, m_2|\ell_3, m_3\rangle$ denotes the 
Clebsch-Gordan coefficients, and $\boldsymbol{\xi}_{-1} = (1,-i,0)^T/\sqrt{2}$, 
$\boldsymbol{\xi}_{0} = (0,0,1)^T$ and $\boldsymbol{\xi}_{1} = 
(-1,-i,0)^T/\sqrt{2}$ are the spherical basis vectors.

\bibliography{sources}

\begin{thebibliography}{43}
\expandafter\ifx\csname natexlab\endcsname\relax\def\natexlab#1{#1}\fi
\expandafter\ifx\csname bibnamefont\endcsname\relax
  \def\bibnamefont#1{#1}\fi
\expandafter\ifx\csname bibfnamefont\endcsname\relax
  \def\bibfnamefont#1{#1}\fi
\expandafter\ifx\csname citenamefont\endcsname\relax
  \def\citenamefont#1{#1}\fi
\expandafter\ifx\csname url\endcsname\relax
  \def\url#1{\texttt{#1}}\fi
\expandafter\ifx\csname urlprefix\endcsname\relax\def\urlprefix{URL }\fi
\providecommand{\bibinfo}[2]{#2}
\providecommand{\eprint}[2][]{\url{#2}}

\bibitem[{\citenamefont{Frenkel}(1931)}]{frenkel1931}
\bibinfo{author}{\bibfnamefont{J.}~\bibnamefont{Frenkel}},
  \bibinfo{journal}{Phys. Rev.} \textbf{\bibinfo{volume}{37}},
  \bibinfo{pages}{17} (\bibinfo{year}{1931}).

\bibitem[{\citenamefont{Wannier}(1937)}]{wannier1937}
\bibinfo{author}{\bibfnamefont{G.~H.} \bibnamefont{Wannier}},
  \bibinfo{journal}{Phys. Rev.} \textbf{\bibinfo{volume}{52}},
  \bibinfo{pages}{191} (\bibinfo{year}{1937}).

\bibitem[{\citenamefont{Gross}(1956)}]{gross1956}
\bibinfo{author}{\bibfnamefont{E.~F.} \bibnamefont{Gross}},
  \bibinfo{journal}{Il Nuovo Cimento (1955-1965)} \textbf{\bibinfo{volume}{3}},
  \bibinfo{pages}{672} (\bibinfo{year}{1956}).

\bibitem[{\citenamefont{Kazimierczuk et~al.}(2014)\citenamefont{Kazimierczuk,
  Fr{\"o}hlich, Scheel, Stolz, and Bayer}}]{kazimierczuk2014}
\bibinfo{author}{\bibfnamefont{T.}~\bibnamefont{Kazimierczuk}},
  \bibinfo{author}{\bibfnamefont{D.}~\bibnamefont{Fr{\"o}hlich}},
  \bibinfo{author}{\bibfnamefont{S.}~\bibnamefont{Scheel}},
  \bibinfo{author}{\bibfnamefont{H.}~\bibnamefont{Stolz}}, \bibnamefont{and}
  \bibinfo{author}{\bibfnamefont{M.}~\bibnamefont{Bayer}},
  \bibinfo{journal}{Nature} \textbf{\bibinfo{volume}{514}},
  \bibinfo{pages}{343} (\bibinfo{year}{2014}).

\bibitem[{\citenamefont{Thewes et~al.}(2015)\citenamefont{Thewes,
  Heck{\"o}tter, Kazimierczuk, A{\ss}mann, Fr{\"o}hlich, Bayer, Semina, and
  Glazov}}]{thewes2015}
\bibinfo{author}{\bibfnamefont{J.}~\bibnamefont{Thewes}},
  \bibinfo{author}{\bibfnamefont{J.}~\bibnamefont{Heck{\"o}tter}},
  \bibinfo{author}{\bibfnamefont{T.}~\bibnamefont{Kazimierczuk}},
  \bibinfo{author}{\bibfnamefont{M.}~\bibnamefont{A{\ss}mann}},
  \bibinfo{author}{\bibfnamefont{D.}~\bibnamefont{Fr{\"o}hlich}},
  \bibinfo{author}{\bibfnamefont{M.}~\bibnamefont{Bayer}},
  \bibinfo{author}{\bibfnamefont{M.~A.} \bibnamefont{Semina}},
  \bibnamefont{and} \bibinfo{author}{\bibfnamefont{M.~M.}
  \bibnamefont{Glazov}}, \bibinfo{journal}{Phys. Rev. Lett.}
  \textbf{\bibinfo{volume}{115}}, \bibinfo{pages}{027402}
  (\bibinfo{year}{2015}).

\bibitem[{\citenamefont{Sch{\"o}ne
  et~al.}(2016{\natexlab{a}})\citenamefont{Sch{\"o}ne, Kr{\"u}ger,
  Gr{\"u}nwald, Stolz, Scheel, A{\ss}mann, Heck{\"o}tter, Thewes, Fr{\"o}hlich,
  and Bayer}}]{schoene2016_prb}
\bibinfo{author}{\bibfnamefont{F.}~\bibnamefont{Sch{\"o}ne}},
  \bibinfo{author}{\bibfnamefont{S.~O.} \bibnamefont{Kr{\"u}ger}},
  \bibinfo{author}{\bibfnamefont{P.}~\bibnamefont{Gr{\"u}nwald}},
  \bibinfo{author}{\bibfnamefont{H.}~\bibnamefont{Stolz}},
  \bibinfo{author}{\bibfnamefont{S.}~\bibnamefont{Scheel}},
  \bibinfo{author}{\bibfnamefont{M.}~\bibnamefont{A{\ss}mann}},
  \bibinfo{author}{\bibfnamefont{J.}~\bibnamefont{Heck{\"o}tter}},
  \bibinfo{author}{\bibfnamefont{J.}~\bibnamefont{Thewes}},
  \bibinfo{author}{\bibfnamefont{D.}~\bibnamefont{Fr{\"o}hlich}},
  \bibnamefont{and} \bibinfo{author}{\bibfnamefont{M.}~\bibnamefont{Bayer}},
  \bibinfo{journal}{Phys. Rev. B} \textbf{\bibinfo{volume}{93}},
  \bibinfo{pages}{075203} (\bibinfo{year}{2016}{\natexlab{a}}).

\bibitem[{\citenamefont{Sch{\"o}ne
  et~al.}(2016{\natexlab{b}})\citenamefont{Sch{\"o}ne, Kr{\"u}ger,
  Gr{\"u}nwald, A{\ss}mann, Heck{\"o}tter, Thewes, Stolz, Fr{\"o}hlich, Bayer,
  and Scheel}}]{schoene2016_jpb}
\bibinfo{author}{\bibfnamefont{F.}~\bibnamefont{Sch{\"o}ne}},
  \bibinfo{author}{\bibfnamefont{S.~O.} \bibnamefont{Kr{\"u}ger}},
  \bibinfo{author}{\bibfnamefont{P.}~\bibnamefont{Gr{\"u}nwald}},
  \bibinfo{author}{\bibfnamefont{M.}~\bibnamefont{A{\ss}mann}},
  \bibinfo{author}{\bibfnamefont{J.}~\bibnamefont{Heck{\"o}tter}},
  \bibinfo{author}{\bibfnamefont{J.}~\bibnamefont{Thewes}},
  \bibinfo{author}{\bibfnamefont{H.}~\bibnamefont{Stolz}},
  \bibinfo{author}{\bibfnamefont{D.}~\bibnamefont{Fr{\"o}hlich}},
  \bibinfo{author}{\bibfnamefont{M.}~\bibnamefont{Bayer}}, \bibnamefont{and}
  \bibinfo{author}{\bibfnamefont{S.}~\bibnamefont{Scheel}},
  \bibinfo{journal}{J. Phys. B: At. Mol. Opt. Phys.}
  \textbf{\bibinfo{volume}{49}}, \bibinfo{pages}{134003}
  (\bibinfo{year}{2016}{\natexlab{b}}).

\bibitem[{\citenamefont{Walther
  et~al.}(2018{\natexlab{a}})\citenamefont{Walther, Kr{\"u}ger, Scheel, and
  Pohl}}]{walther2018}
\bibinfo{author}{\bibfnamefont{V.}~\bibnamefont{Walther}},
  \bibinfo{author}{\bibfnamefont{S.~O.} \bibnamefont{Kr{\"u}ger}},
  \bibinfo{author}{\bibfnamefont{S.}~\bibnamefont{Scheel}}, \bibnamefont{and}
  \bibinfo{author}{\bibfnamefont{T.}~\bibnamefont{Pohl}},
  \bibinfo{journal}{Phys. Rev. B} \textbf{\bibinfo{volume}{98}},
  \bibinfo{pages}{165201} (\bibinfo{year}{2018}{\natexlab{a}}).

\bibitem[{\citenamefont{Heck{\"o}tter et~al.}(2018)\citenamefont{Heck{\"o}tter,
  Freitag, Fr{\"o}hlich, A{\ss}mann, Bayer, Gr{\"u}nwald, Sch{\"o}ne, Semkat,
  Stolz, and Scheel}}]{heckotter2018}
\bibinfo{author}{\bibfnamefont{J.}~\bibnamefont{Heck{\"o}tter}},
  \bibinfo{author}{\bibfnamefont{M.}~\bibnamefont{Freitag}},
  \bibinfo{author}{\bibfnamefont{D.}~\bibnamefont{Fr{\"o}hlich}},
  \bibinfo{author}{\bibfnamefont{M.}~\bibnamefont{A{\ss}mann}},
  \bibinfo{author}{\bibfnamefont{M.}~\bibnamefont{Bayer}},
  \bibinfo{author}{\bibfnamefont{P.}~\bibnamefont{Gr{\"u}nwald}},
  \bibinfo{author}{\bibfnamefont{F.}~\bibnamefont{Sch{\"o}ne}},
  \bibinfo{author}{\bibfnamefont{D.}~\bibnamefont{Semkat}},
  \bibinfo{author}{\bibfnamefont{H.}~\bibnamefont{Stolz}}, \bibnamefont{and}
  \bibinfo{author}{\bibfnamefont{S.}~\bibnamefont{Scheel}},
  \bibinfo{journal}{Phys. Rev. Lett.} \textbf{\bibinfo{volume}{121}},
  \bibinfo{pages}{097401} (\bibinfo{year}{2018}).

\bibitem[{\citenamefont{Schweiner et~al.}(2016)\citenamefont{Schweiner, Main,
  Feldmaier, Wunner, and Uihlein}}]{schweiner2016impact}
\bibinfo{author}{\bibfnamefont{F.}~\bibnamefont{Schweiner}},
  \bibinfo{author}{\bibfnamefont{J.}~\bibnamefont{Main}},
  \bibinfo{author}{\bibfnamefont{M.}~\bibnamefont{Feldmaier}},
  \bibinfo{author}{\bibfnamefont{G.}~\bibnamefont{Wunner}}, \bibnamefont{and}
  \bibinfo{author}{\bibfnamefont{C.}~\bibnamefont{Uihlein}},
  \bibinfo{journal}{Phys. Rev. B} \textbf{\bibinfo{volume}{93}},
  \bibinfo{pages}{195203} (\bibinfo{year}{2016}).

\bibitem[{\citenamefont{Stolz et~al.}(2018)\citenamefont{Stolz, Sch{\"o}ne, and
  Semkat}}]{stolz2018}
\bibinfo{author}{\bibfnamefont{H.}~\bibnamefont{Stolz}},
  \bibinfo{author}{\bibfnamefont{F.}~\bibnamefont{Sch{\"o}ne}},
  \bibnamefont{and} \bibinfo{author}{\bibfnamefont{D.}~\bibnamefont{Semkat}},
  \bibinfo{journal}{New J. Phys.} \textbf{\bibinfo{volume}{20}},
  \bibinfo{pages}{023019} (\bibinfo{year}{2018}).

\bibitem[{\citenamefont{Schweiner
  et~al.}(2017{\natexlab{a}})\citenamefont{Schweiner, Main, Wunner, Freitag,
  Heck{\"o}tter, Uihlein, A{\ss}mann, Fr{\"o}hlich, and
  Bayer}}]{schweiner2017magnetoexcitons}
\bibinfo{author}{\bibfnamefont{F.}~\bibnamefont{Schweiner}},
  \bibinfo{author}{\bibfnamefont{J.}~\bibnamefont{Main}},
  \bibinfo{author}{\bibfnamefont{G.}~\bibnamefont{Wunner}},
  \bibinfo{author}{\bibfnamefont{M.}~\bibnamefont{Freitag}},
  \bibinfo{author}{\bibfnamefont{J.}~\bibnamefont{Heck{\"o}tter}},
  \bibinfo{author}{\bibfnamefont{C.}~\bibnamefont{Uihlein}},
  \bibinfo{author}{\bibfnamefont{M.}~\bibnamefont{A{\ss}mann}},
  \bibinfo{author}{\bibfnamefont{D.}~\bibnamefont{Fr{\"o}hlich}},
  \bibnamefont{and} \bibinfo{author}{\bibfnamefont{M.}~\bibnamefont{Bayer}},
  \bibinfo{journal}{Phys. Rev. B} \textbf{\bibinfo{volume}{95}},
  \bibinfo{pages}{035202} (\bibinfo{year}{2017}{\natexlab{a}}).

\bibitem[{\citenamefont{Kurz et~al.}(2017)\citenamefont{Kurz, Gr{\"u}nwald, and
  Scheel}}]{kurz2017}
\bibinfo{author}{\bibfnamefont{M.}~\bibnamefont{Kurz}},
  \bibinfo{author}{\bibfnamefont{P.}~\bibnamefont{Gr{\"u}nwald}},
  \bibnamefont{and} \bibinfo{author}{\bibfnamefont{S.}~\bibnamefont{Scheel}},
  \bibinfo{journal}{Phys. Rev. B} \textbf{\bibinfo{volume}{95}},
  \bibinfo{pages}{245205} (\bibinfo{year}{2017}).

\bibitem[{\citenamefont{A{\ss}mann et~al.}(2016)\citenamefont{A{\ss}mann,
  Thewes, Fr{\"o}hlich, and Bayer}}]{assmann2016}
\bibinfo{author}{\bibfnamefont{M.}~\bibnamefont{A{\ss}mann}},
  \bibinfo{author}{\bibfnamefont{J.}~\bibnamefont{Thewes}},
  \bibinfo{author}{\bibfnamefont{D.}~\bibnamefont{Fr{\"o}hlich}},
  \bibnamefont{and} \bibinfo{author}{\bibfnamefont{M.}~\bibnamefont{Bayer}},
  \bibinfo{journal}{Nature Materials} \textbf{\bibinfo{volume}{15}},
  \bibinfo{pages}{741} (\bibinfo{year}{2016}).

\bibitem[{\citenamefont{Schweiner
  et~al.}(2017{\natexlab{b}})\citenamefont{Schweiner, Main, and
  Wunner}}]{schweiner2017magnetoexcitons2}
\bibinfo{author}{\bibfnamefont{F.}~\bibnamefont{Schweiner}},
  \bibinfo{author}{\bibfnamefont{J.}~\bibnamefont{Main}}, \bibnamefont{and}
  \bibinfo{author}{\bibfnamefont{G.}~\bibnamefont{Wunner}},
  \bibinfo{journal}{Phys. Rev. Lett.} \textbf{\bibinfo{volume}{118}},
  \bibinfo{pages}{046401} (\bibinfo{year}{2017}{\natexlab{b}}).

\bibitem[{\citenamefont{Nikitine}(1959)}]{nikitine1959}
\bibinfo{author}{\bibfnamefont{S.}~\bibnamefont{Nikitine}},
  \bibinfo{journal}{Philosophical Magazine} \textbf{\bibinfo{volume}{4}},
  \bibinfo{pages}{1} (\bibinfo{year}{1959}).

\bibitem[{\citenamefont{Gross}(1962)}]{gross1962}
\bibinfo{author}{\bibfnamefont{E.}~\bibnamefont{Gross}},
  \bibinfo{journal}{Soviet Physics Uspekhi} \textbf{\bibinfo{volume}{5}},
  \bibinfo{pages}{195} (\bibinfo{year}{1962}).

\bibitem[{\citenamefont{Schmutzler et~al.}(2013)\citenamefont{Schmutzler,
  Fr{\"o}hlich, and Bayer}}]{schmutzler2013}
\bibinfo{author}{\bibfnamefont{J.}~\bibnamefont{Schmutzler}},
  \bibinfo{author}{\bibfnamefont{D.}~\bibnamefont{Fr{\"o}hlich}},
  \bibnamefont{and} \bibinfo{author}{\bibfnamefont{M.}~\bibnamefont{Bayer}},
  \bibinfo{journal}{Phys. Rev. B} \textbf{\bibinfo{volume}{87}},
  \bibinfo{pages}{245202} (\bibinfo{year}{2013}).

\bibitem[{\citenamefont{J{\"o}rger et~al.}(2003)\citenamefont{J{\"o}rger,
  Tsitsishvili, Fleck, and Klingshirn}}]{jorger2003}
\bibinfo{author}{\bibfnamefont{M.}~\bibnamefont{J{\"o}rger}},
  \bibinfo{author}{\bibfnamefont{E.}~\bibnamefont{Tsitsishvili}},
  \bibinfo{author}{\bibfnamefont{T.}~\bibnamefont{Fleck}}, \bibnamefont{and}
  \bibinfo{author}{\bibfnamefont{C.}~\bibnamefont{Klingshirn}},
  \bibinfo{journal}{phys. stat. sol. (b)} \textbf{\bibinfo{volume}{238}},
  \bibinfo{pages}{470} (\bibinfo{year}{2003}).

\bibitem[{\citenamefont{Ziemkiewicz and
  Zieli{\'n}ska-Raczy{\'n}ska}(2018)}]{ziemkiewicz2018}
\bibinfo{author}{\bibfnamefont{D.}~\bibnamefont{Ziemkiewicz}} \bibnamefont{and}
  \bibinfo{author}{\bibfnamefont{S.}~\bibnamefont{Zieli{\'n}ska-Raczy{\'n}ska}},
  \bibinfo{journal}{Opt. Lett.} \textbf{\bibinfo{volume}{43}},
  \bibinfo{pages}{3742} (\bibinfo{year}{2018}).

\bibitem[{\citenamefont{Ziemkiewicz and
  Zieli{\'n}ska-Raczy{\'n}ska}(2019)}]{ziemkiewicz2019}
\bibinfo{author}{\bibfnamefont{D.}~\bibnamefont{Ziemkiewicz}} \bibnamefont{and}
  \bibinfo{author}{\bibfnamefont{S.}~\bibnamefont{Zieli{\'n}ska-Raczy{\'n}ska}},
  \bibinfo{journal}{Opt. Exp.} \textbf{\bibinfo{volume}{27}},
  \bibinfo{pages}{16983} (\bibinfo{year}{2019}).

\bibitem[{\citenamefont{Takahata and Naka}(2018)}]{takahata2018.1}
\bibinfo{author}{\bibfnamefont{M.}~\bibnamefont{Takahata}} \bibnamefont{and}
  \bibinfo{author}{\bibfnamefont{N.}~\bibnamefont{Naka}},
  \bibinfo{journal}{Phys. Rev. B} \textbf{\bibinfo{volume}{98}},
  \bibinfo{pages}{195205} (\bibinfo{year}{2018}).

\bibitem[{\citenamefont{Takahata et~al.}(2018)\citenamefont{Takahata, Tanaka,
  and Naka}}]{takahata2018.2}
\bibinfo{author}{\bibfnamefont{M.}~\bibnamefont{Takahata}},
  \bibinfo{author}{\bibfnamefont{K.}~\bibnamefont{Tanaka}}, \bibnamefont{and}
  \bibinfo{author}{\bibfnamefont{N.}~\bibnamefont{Naka}},
  \bibinfo{journal}{Phys. Rev. Lett.} \textbf{\bibinfo{volume}{121}},
  \bibinfo{pages}{173604} (\bibinfo{year}{2018}).

\bibitem[{\citenamefont{Gr{\"u}nwald et~al.}(2016)\citenamefont{Gr{\"u}nwald,
  A{\ss}mann, Heck{\"o}tter, Fr{\"o}hlich, Bayer, Stolz, and
  Scheel}}]{grunwald2016}
\bibinfo{author}{\bibfnamefont{P.}~\bibnamefont{Gr{\"u}nwald}},
  \bibinfo{author}{\bibfnamefont{M.}~\bibnamefont{A{\ss}mann}},
  \bibinfo{author}{\bibfnamefont{J.}~\bibnamefont{Heck{\"o}tter}},
  \bibinfo{author}{\bibfnamefont{D.}~\bibnamefont{Fr{\"o}hlich}},
  \bibinfo{author}{\bibfnamefont{M.}~\bibnamefont{Bayer}},
  \bibinfo{author}{\bibfnamefont{H.}~\bibnamefont{Stolz}}, \bibnamefont{and}
  \bibinfo{author}{\bibfnamefont{S.}~\bibnamefont{Scheel}},
  \bibinfo{journal}{Phys. Rev. Lett.} \textbf{\bibinfo{volume}{117}},
  \bibinfo{pages}{133003} (\bibinfo{year}{2016}).

\bibitem[{\citenamefont{Khazali et~al.}(2017)\citenamefont{Khazali, Heshami,
  and Simon}}]{khazali2017}
\bibinfo{author}{\bibfnamefont{M.}~\bibnamefont{Khazali}},
  \bibinfo{author}{\bibfnamefont{K.}~\bibnamefont{Heshami}}, \bibnamefont{and}
  \bibinfo{author}{\bibfnamefont{C.}~\bibnamefont{Simon}}, \bibinfo{journal}{J.
  Phys. B: At. Mol. Opt. Phys.} \textbf{\bibinfo{volume}{50}},
  \bibinfo{pages}{215301} (\bibinfo{year}{2017}).

\bibitem[{\citenamefont{Walther
  et~al.}(2018{\natexlab{b}})\citenamefont{Walther, Johne, and
  Pohl}}]{walther2018b}
\bibinfo{author}{\bibfnamefont{V.}~\bibnamefont{Walther}},
  \bibinfo{author}{\bibfnamefont{R.}~\bibnamefont{Johne}}, \bibnamefont{and}
  \bibinfo{author}{\bibfnamefont{T.}~\bibnamefont{Pohl}},
  \bibinfo{journal}{Nature Communications} \textbf{\bibinfo{volume}{9}},
  \bibinfo{pages}{1309} (\bibinfo{year}{2018}{\natexlab{b}}).

\bibitem[{\citenamefont{Werner and Hochheimer}(1982)}]{werner1982}
\bibinfo{author}{\bibfnamefont{A.}~\bibnamefont{Werner}} \bibnamefont{and}
  \bibinfo{author}{\bibfnamefont{H.~D.} \bibnamefont{Hochheimer}},
  \bibinfo{journal}{Phys. Rev. B} \textbf{\bibinfo{volume}{25}},
  \bibinfo{pages}{5929} (\bibinfo{year}{1982}).

\bibitem[{\citenamefont{Carabatos et~al.}(1968)\citenamefont{Carabatos,
  Diffin{\'e}, and Sieskind}}]{carabatos1968}
\bibinfo{author}{\bibfnamefont{C.}~\bibnamefont{Carabatos}},
  \bibinfo{author}{\bibfnamefont{A.}~\bibnamefont{Diffin{\'e}}},
  \bibnamefont{and} \bibinfo{author}{\bibfnamefont{M.}~\bibnamefont{Sieskind}},
  \bibinfo{journal}{Journal de Physique} \textbf{\bibinfo{volume}{29}},
  \bibinfo{pages}{529} (\bibinfo{year}{1968}).

\bibitem[{\citenamefont{Kavoulakis et~al.}(1997)\citenamefont{Kavoulakis,
  Chang, and Baym}}]{kavoulakis1997}
\bibinfo{author}{\bibfnamefont{G.~M.} \bibnamefont{Kavoulakis}},
  \bibinfo{author}{\bibfnamefont{Y.-C.} \bibnamefont{Chang}}, \bibnamefont{and}
  \bibinfo{author}{\bibfnamefont{G.}~\bibnamefont{Baym}},
  \bibinfo{journal}{Phys. Rev. B} \textbf{\bibinfo{volume}{55}},
  \bibinfo{pages}{7593} (\bibinfo{year}{1997}).

\bibitem[{\citenamefont{Itoh and Narita}(1975)}]{itoh1975}
\bibinfo{author}{\bibfnamefont{T.}~\bibnamefont{Itoh}} \bibnamefont{and}
  \bibinfo{author}{\bibfnamefont{S.-i.} \bibnamefont{Narita}},
  \bibinfo{journal}{J. Phys. Soc. Jpn.} \textbf{\bibinfo{volume}{39}},
  \bibinfo{pages}{140} (\bibinfo{year}{1975}).

\bibitem[{\citenamefont{Waters et~al.}(1980)\citenamefont{Waters, Pollak,
  Bruce, and Cummins}}]{Waters1980}
\bibinfo{author}{\bibfnamefont{R.~G.} \bibnamefont{Waters}},
  \bibinfo{author}{\bibfnamefont{F.~H.} \bibnamefont{Pollak}},
  \bibinfo{author}{\bibfnamefont{R.~H.} \bibnamefont{Bruce}}, \bibnamefont{and}
  \bibinfo{author}{\bibfnamefont{H.~Z.} \bibnamefont{Cummins}},
  \bibinfo{journal}{Phys. Rev. B} \textbf{\bibinfo{volume}{21}},
  \bibinfo{pages}{1665} (\bibinfo{year}{1980}).

\bibitem[{\citenamefont{Suzuki and Hensel}(1974)}]{suzuki1974}
\bibinfo{author}{\bibfnamefont{K.}~\bibnamefont{Suzuki}} \bibnamefont{and}
  \bibinfo{author}{\bibfnamefont{J.~C.} \bibnamefont{Hensel}},
  \bibinfo{journal}{Phys. Rev. B} \textbf{\bibinfo{volume}{9}},
  \bibinfo{pages}{4184} (\bibinfo{year}{1974}).

\bibitem[{\citenamefont{French et~al.}(2008)\citenamefont{French, Schwartz,
  Stolz, and Redmer}}]{french2008}
\bibinfo{author}{\bibfnamefont{M.}~\bibnamefont{French}},
  \bibinfo{author}{\bibfnamefont{R.}~\bibnamefont{Schwartz}},
  \bibinfo{author}{\bibfnamefont{H.}~\bibnamefont{Stolz}}, \bibnamefont{and}
  \bibinfo{author}{\bibfnamefont{R.}~\bibnamefont{Redmer}},
  \bibinfo{journal}{J. Phys.: Condens. Matter} \textbf{\bibinfo{volume}{21}},
  \bibinfo{pages}{015502} (\bibinfo{year}{2008}).

\bibitem[{\citenamefont{Hodby et~al.}(1976)\citenamefont{Hodby, Jenkins,
  Schwab, Tamura, and Trivich}}]{hodby1976}
\bibinfo{author}{\bibfnamefont{J.~W.} \bibnamefont{Hodby}},
  \bibinfo{author}{\bibfnamefont{T.~E.} \bibnamefont{Jenkins}},
  \bibinfo{author}{\bibfnamefont{C.}~\bibnamefont{Schwab}},
  \bibinfo{author}{\bibfnamefont{H.}~\bibnamefont{Tamura}}, \bibnamefont{and}
  \bibinfo{author}{\bibfnamefont{D.}~\bibnamefont{Trivich}},
  \bibinfo{journal}{J. Phys. C: Solid State Phys.}
  \textbf{\bibinfo{volume}{9}}, \bibinfo{pages}{1429} (\bibinfo{year}{1976}).

\bibitem[{\citenamefont{Schweiner
  et~al.}(2017{\natexlab{c}})\citenamefont{Schweiner, Main, Wunner, and
  Uihlein}}]{schweiner2017even}
\bibinfo{author}{\bibfnamefont{F.}~\bibnamefont{Schweiner}},
  \bibinfo{author}{\bibfnamefont{J.}~\bibnamefont{Main}},
  \bibinfo{author}{\bibfnamefont{G.}~\bibnamefont{Wunner}}, \bibnamefont{and}
  \bibinfo{author}{\bibfnamefont{C.}~\bibnamefont{Uihlein}},
  \bibinfo{journal}{Phys. Rev. B} \textbf{\bibinfo{volume}{95}},
  \bibinfo{pages}{195201} (\bibinfo{year}{2017}{\natexlab{c}}).

\bibitem[{\citenamefont{Alvermann and Fehske}(2018)}]{alvermann2018}
\bibinfo{author}{\bibfnamefont{A.}~\bibnamefont{Alvermann}} \bibnamefont{and}
  \bibinfo{author}{\bibfnamefont{H.}~\bibnamefont{Fehske}},
  \bibinfo{journal}{J. Phys. B: At. Mol. Opt. Phys.}
  \textbf{\bibinfo{volume}{51}}, \bibinfo{pages}{044001}
  (\bibinfo{year}{2018}).

\bibitem[{\citenamefont{Konzelmann et~al.}(2019)\citenamefont{Konzelmann,
  Kr{\"u}ger, and Giessen}}]{konzelmann2019}
\bibinfo{author}{\bibfnamefont{A.~M.} \bibnamefont{Konzelmann}},
  \bibinfo{author}{\bibfnamefont{S.~O.} \bibnamefont{Kr{\"u}ger}},
  \bibnamefont{and} \bibinfo{author}{\bibfnamefont{H.}~\bibnamefont{Giessen}},
  \emph{\bibinfo{title}{Interaction of {OAM} light with {R}ydberg excitons:
  Modifying dipole selection rules}}, \bibinfo{howpublished}{arXiv:1905.07131}
  (\bibinfo{year}{2019}).

\bibitem[{\citenamefont{Schmiegelow et~al.}(2016)\citenamefont{Schmiegelow,
  Schulz, Kaufmann, Ruster, Poschinger, and Schmidt-Kaler}}]{schmiegelow2016}
\bibinfo{author}{\bibfnamefont{C.~T.} \bibnamefont{Schmiegelow}},
  \bibinfo{author}{\bibfnamefont{J.}~\bibnamefont{Schulz}},
  \bibinfo{author}{\bibfnamefont{H.}~\bibnamefont{Kaufmann}},
  \bibinfo{author}{\bibfnamefont{T.}~\bibnamefont{Ruster}},
  \bibinfo{author}{\bibfnamefont{U.~G.} \bibnamefont{Poschinger}},
  \bibnamefont{and}
  \bibinfo{author}{\bibfnamefont{F.}~\bibnamefont{Schmidt-Kaler}},
  \bibinfo{journal}{Nature Communications} \textbf{\bibinfo{volume}{7}},
  \bibinfo{pages}{12998} (\bibinfo{year}{2016}).

\bibitem[{\citenamefont{Afanasev et~al.}(2018)\citenamefont{Afanasev, Carlson,
  Schmiegelow, Schulz, Schmidt-Kaler, and Solyanik}}]{afanasev2018}
\bibinfo{author}{\bibfnamefont{A.}~\bibnamefont{Afanasev}},
  \bibinfo{author}{\bibfnamefont{C.~E.} \bibnamefont{Carlson}},
  \bibinfo{author}{\bibfnamefont{C.~T.} \bibnamefont{Schmiegelow}},
  \bibinfo{author}{\bibfnamefont{J.}~\bibnamefont{Schulz}},
  \bibinfo{author}{\bibfnamefont{F.}~\bibnamefont{Schmidt-Kaler}},
  \bibnamefont{and} \bibinfo{author}{\bibfnamefont{M.}~\bibnamefont{Solyanik}},
  \bibinfo{journal}{New J. Phys.} \textbf{\bibinfo{volume}{20}},
  \bibinfo{pages}{023032} (\bibinfo{year}{2018}).

\bibitem[{\citenamefont{Voon and Willatzen}(2009)}]{kp-voon}
\bibinfo{author}{\bibfnamefont{L.~C. L.~Y.} \bibnamefont{Voon}}
  \bibnamefont{and}
  \bibinfo{author}{\bibfnamefont{M.}~\bibnamefont{Willatzen}},
  \emph{\bibinfo{title}{The $k\cdot p$ Method}} (\bibinfo{publisher}{Springer
  Verlag, Berlin, Heidelberg}, \bibinfo{year}{2009}).

\bibitem[{\citenamefont{Luttinger}(1956)}]{luttinger1956}
\bibinfo{author}{\bibfnamefont{J.}~\bibnamefont{Luttinger}},
  \bibinfo{journal}{Phys. Rev.} \textbf{\bibinfo{volume}{102}},
  \bibinfo{pages}{1030} (\bibinfo{year}{1956}).

\bibitem[{\citenamefont{Koster et~al.}(1963)\citenamefont{Koster, Dimmock,
  Wheeler, and Statz}}]{koster1963}
\bibinfo{author}{\bibfnamefont{G.~F.} \bibnamefont{Koster}},
  \bibinfo{author}{\bibfnamefont{J.~O.} \bibnamefont{Dimmock}},
  \bibinfo{author}{\bibfnamefont{R.~G.} \bibnamefont{Wheeler}},
  \bibnamefont{and} \bibinfo{author}{\bibfnamefont{H.}~\bibnamefont{Statz}},
  \bibinfo{journal}{Properties of the 32-point groups}  (\bibinfo{year}{1963}).

\bibitem[{\citenamefont{Von~der Lage and Bethe}(1947)}]{vdLage1947}
\bibinfo{author}{\bibfnamefont{F.~C.} \bibnamefont{Von~der Lage}}
  \bibnamefont{and} \bibinfo{author}{\bibfnamefont{H.~A.} \bibnamefont{Bethe}},
  \bibinfo{journal}{Phys. Rev.} \textbf{\bibinfo{volume}{71}},
  \bibinfo{pages}{612} (\bibinfo{year}{1947}).

\end{thebibliography}
\end{document}